\documentclass[aps,showpacs,pre,amssymb]{revtex4}
\usepackage{floatflt}
\usepackage{hhline}
\usepackage[dvips]{graphicx}
\newfont{\myBbb}{msbm10 scaled 1200}
\newcommand{\la}{\langle}
\newcommand{\ra}{\rangle}
\newcommand{\mod}[1]{\ (\bmod\ #1)}
\newcommand{\N}{{\mbox{\myBbb N}}}
\newcommand{\Z}{{\mbox{\myBbb Z}}}
\begin{document}
\title{Periodic orbits of the ensemble of Sinai-Arnold
cat maps and pseudorandom number generation}
\author{L.\ Barash\,$^{1,a)}$ and L.N.\ Shchur\,$^{1,2,b)}$}
\affiliation{
$^{1)}$Landau Institute for Theoretical Physics, 142432
Chernogolovka, Russia \\
$^{2)}$Materials Science Division, Argonne National Laboratory,
Argonne, Illinois 60439, USA \\
e-mail: \tt $^{a)}$barash@itp.ac.ru, $^{b)}$lev@landau.ac.ru}
\begin{abstract}

We propose methods for constructing high-quality pseudorandom number
generators (RNGs) based on an ensemble of hyperbolic automorphisms of the
unit two-dimensional torus (Sinai--Arnold map or cat map) while keeping a
part of the information hidden. The single cat map provides the random
properties expected from a good RNG and is hence an appropriate building
block for an RNG, although unnecessary correlations are always present in
practice. We show that introducing hidden variables and introducing
rotation in the RNG output, accompanied with the proper initialization,
dramatically suppress these correlations. We analyze the mechanisms of the
single-cat-map correlations analytically and show how to diminish them. We
generalize the Percival--Vivaldi theory in the case of the ensemble of
maps, find the period of the proposed RNG analytically, and also analyze
its properties. We present efficient practical realizations for the RNGs
and check our predictions numerically. We also test our RNGs using the
known stringent batteries of statistical tests and find that the
statistical properties of our best generators are not worse than those of
other best modern generators.

\end{abstract}

\pacs{02.50.Ng, 02.70.Uu, 05.45.-a}
\maketitle

\section{Introduction}
\label{IntroductionSec}

Molecular dynamics and Monte Carlo simulations are important
computational techniques in many areas of science: in quantum
physics~\cite{Beach}, statistical physics~\cite{Landau2000}, nuclear
physics~\cite{Pieper}, quantum chemistry~\cite{Luechow}, material
science~\cite{Bizzari}, among many others. The simulations rely
heavily on the use of random numbers, which are generated by
deterministic recursive rules. Such rules produce pseudorandom
numbers, and it is a great challenge to design random number
generators (RNGs) that behave as realizations of independent
uniformly distributed random variables and approximate ``true
randomness''~\cite{Knuth}.

There are several requirements for a good RNG and its implementation
in a subroutine library. Among them are statistical robustness
(uniform distribution of values at the output with no apparent
correlations), unpredictability, long period, efficiency,
theoretical support (precise prediction of the important properties),
portability and others~\cite{Knuth,Lecuyer,Brent}.

A number of RNGs introduced in the last five decades fulfill most of the requirements
and are successfully used in simulations. Nevertheless, each of them
has some weak properties which may (or may not) influence the results.

The most widely used RNGs can be divided into two classes. The first
class is represented by the Linear Congruential Generator (LCG),
and the second, by Shift Register (SR) generator.

{\it Linear Congruential Generators} (LCGs) are the best-known and (still)
most widely available RNGs in use today. An example of the
realization of an LCG generator is the UNIX {\tt rand} generator
$y_n=(1103515245\; y_{n-1}+12345)\mod{2^{31}}$.
The practical recommendation is that LCGs should be avoided
for applications dealing with the geometric behavior of random
vectors in high dimensions because of the bad geometric structure of
the vectors that they produce~\cite{Knuth,Coveyou}.

{\it Generalized Feedback Shift Register} (GFSR) sequences are widely
used in many areas of computational and simulational physics. These
RNGs are quite fast and possess huge periods given a proper choice
of the underlying primitive trinomials~\cite{Golomb}. This makes them
particularly well suited for applications that require many
pseudorandom numbers. But several flaws have been observed
in the statistical properties of these generators, which can result in
systematic errors in Monte Carlo simulations. Typical examples include
the Wolff single cluster algorithm for the 2D Ising model
simulation~\cite{SR1}, random and self-avoiding walks~\cite{Grass}, and the 3D
Blume--Capel model using local Metropolis updating~\cite{SR3}.

Modern modifications and generalizations to the LCG and GFSR methods
have much better periodic and statistical properties. Some examples
are the Mersenne twister~\cite{MT} (this generator employs the modified
and generalized GFSR scheme), combined LCGs generators~\cite{CombinedLCG}
and combined Tausworthe generators~\cite{CombinedTausw,LFSR113}.

Most RNGs used today can be easily deciphered. Perhaps the
generator with the best unpredictability properties known today is
the BBS generator~\cite{BBS, BBSImpr}, which is proved to be
polynomial-time perfect under certain reasonable
assumptions~\cite{BBS,Lecuyer} if the size $s$ of the generator is
sufficiently large. This generator is rather slow for practical use
because its speed decreases rapidly as $s$ increases.
The discussion of cryptographic RNG is beyond our analysis.

We propose using an ensemble of simple nonlinear dynamical systems to
construct an RNG. Of course, not all dynamical systems are useful. For
instance, baker's transformation is a simple example of a chaotic
system: it is area preserving and deterministic, and its state is maintained
in a bounded domain. The base of baker's transformation is the
Bernoulli shift $x_{n+1}=2x_n\mod{1}$, it yields a sequence of random
numbers provided we have a random irrational seed. But in real
computation, the seed number has finite complexity, and the number of
available bits decreases at each step. Obviously, there is no practical use
of this scheme for an RNG.

The logistic map~\cite{Licht,Schuster} also does not help to construct
an RNG. First, manipulation with real values of fixed accuracy leads to
significant errors during long orbits. Second, the sequence of
numbers generated by a logistic map does not have a uniform
distribution~\cite{Schuster}. Also, the logistic map represents a
chaotic dynamical system only for isolated values of a parameter. Even
small deviations from these isolated values lead to creating
subregions in the phase space, i.e., the orbit of the point does not
span the whole phase space.

The next class of dynamical systems is Anosov diffeomorphisms of the
two-dimensional torus, which have attracted much attention in the
context of ergodic theory. Anosov systems have the following
stochastic properties: ergodicity, mixing, sensitive dependence on
initial conditions (which follows from the positivity of the Lyapunov
exponent), and local divergence of all trajectories (which follows from
the positivity of the Kolmogorov--Sinai entropy). These properties resemble
certain properties of randomness. Every Anosov diffeomorphism of the
torus is topologically conjugate to a hyperbolic automorphism, which
can be viewed as a completely chaotic Hamiltonian dynamical system.
Hyperbolic automorphisms are represented by $2{\times}2$-matrixes with
integer entries, a unit determinant, and real eigenvalues, and are known
as {\it cat maps} (there are two reasons for this terminology: first,
CAT is an acronym for Continuous Automorphism of the Torus; second, the
chaotic behavior of these maps is traditionally described by showing the
result of their action on the face of the cat~\cite{ArnoldAvez}). We note
that cat maps are Hamiltonian systems. Indeed, if $k=\mbox{Tr} (M)=
m_{11}+m_{22}>2$, then the action of map (\ref{MatrixM})
on the vector ${p\choose q}$ can be described as the motion
in the phase space specified by the Hamiltonian~\cite{Keating}
$H(p,q)=(k^2-4)^{-1/2}\sinh^{-1}((k^2-4)^{1/2}/2)
(m_{12}p^2-m_{21}q^2+(m_{11}-m_{22})pq)$. Here, $p$ and $q$ are taken
modulo $1$ at each observation (i.e., we preserve only the fractional part
of $p$ and $q$; the integer part is ignored), and observations occur at
integer points of time.

In this paper, we present RNGs based on an ensemble of cat maps and
analyze the requirements for a good RNG with respect to our
scheme. The basic idea is to apply the cat map to a discrete set
of points (there are two modifications: for $g=2^m$ and for prime
$g$, where $(g{\times}g)$ is the lattice) such that each
point belongs to a different periodic trajectory.

A similar utilization of cat maps for an RNG is called the matrix generator
for pseudorandom numbers. It was introduced in~\cite{Grothe,Niederr1}
and discussed for prime values of $g$.
But because the single matrix generator is a generalization of the
linear congruential method, it suffers from both the defects
of LCG~\cite{AfferbachGrothe} and the defects of GFSR (see
Sec.~\ref{RandWalksSec}). The periodic and statistical properties of the
matrix generators and of the equivalent multiple recursive generators
have been studied~\cite{NeiderrSerial,Lecuyer98}, but the single
$2{\times}2$-matrix generator still has significant correlations between
values at the output.

Also, there is an impressive theoretical basis for
relating properties of the periodic orbits of cat maps and
properties of algebraic numbers~\cite{PercivalVivaldi}, which to the
best of our knowledge has never been directly applied to RNG theory.
Applying the ensemble of matrix transformations of the two-dimensional
torus while using only a single bit from the point of each map,
and utilizing rotation in the RNG output are the distinctive features
of our generator.
Also, as for other generators, a proper initialization of the
initial state is important.
As will be seen, the proposed scheme has several advantages. First,
it can essentially reduce correlations and lead to creating an RNG
not worse than other modern RNGs.
Second, both the properties of periodic orbits and the statistical
properties of such a generator can be analyzed both theoretically and
empirically. Several examples of RNGs made by this method,
as well as the effective realizations, are presented.

The generator is introduced in Sec.~\ref{GeneratorSec}. In
Sec.~\ref{StatTestsSec}, we present the results for stringent
statistical tests. Correlations for a single cat map are also
analyzed thoroughly, and some correlations are found by the random
walks test. We analyze the mechanism of these correlations in
Sec.~\ref{RandWalksSec}; they appear to be associated with the
geometric properties of the cat map. We find these correlations
analytically (Sec.~\ref{RandWalksSec}). We provide a method for
obtaining quantities such as the periods of cat maps, the number
of orbits with a given period, and the area in the phase space
swept by the orbits with a given period
(Appendix~\ref{CatMapsPeriod}). We also provide a method for
obtaining periods of the generator for arbitrary parameters of the
map and lattice (Appendix~\ref{RNGPeriod}). This gives the primary
theoretical support of the generator. In particular, we find that
the typical period of the generator for the $2^m{\times}2^m$
lattice is $T_m=3\cdot 2^{m-2}$. The method is based on the work
of Percival and Vivaldi, who transformed the study of the periodic
orbits of cat maps into the modular arithmetic in domains of
quadratic integers. The key ideas needed for our consideration are
briefly reviewed in Appendix~\ref{CatMapsPeriod}.
Appendix~\ref{NormDiscussion} gives the method for analyzing
correlations between orbits of different points and choosing the
proper initial conditions to minimize the correlations.
Appendix~\ref{ThProof} and supplemental
details~\cite{SpaceProofCite} support the other sections, giving
detailed proofs of the underlying results.
Appendix~\ref{RealizationSec} presents the efficient realizations
for several versions of RNG, the initialization techniques, and the
analysis of the speed of the RNGs.

\section{The Generator}
\label{GeneratorSec}
\subsection{Description of the method}

We consider hyperbolic automorphisms of the unit two-dimensional torus
(the square $(0,1]\times(0,1]$ with the opposite sides identified).
The action of a given cat map $R$ is defined as follows: first,
we transform the phase space by the matrix

\begin{equation}
M=\left(\begin{array}{cc} m_{11}&m_{12}\\m_{21}&m_{22}\\
\end{array}\right)
\in SL_2(\Z);
\label{MatrixM}
\end{equation}

\noindent second, we take the fractional parts in $(0,1)$ of both
coordinates. Here $SL_2(\Z)$ denotes the special linear group of
degree $2$ over the ring of integers, i.e., the elements of $M$ are integers,
$\det M=1$, and the eigenvalues of $M$ are $\lambda=(k\pm\sqrt{k^2-4})/2$,
where $k={\rm Tr}(M)$ is the trace of the matrix $M$. The eigenvalues
should be real because complex values of $\lambda$ lead to a nonergodic
dynamical process, and the hyperbolicity condition is $|k|>2$.

It is easy to prove that the periodic orbits of the hyperbolic toral
automorphism $R$ consist precisely of those points that have rational
coordinates~\cite{ArnoldAvez,PercivalVivaldi,Keating}. Hence, it is
natural to consider the dynamics of the map defined on the set of
points with rational coordinates that share a given denominator $g$.
The lattice of such points is invariant under the action of the
cat maps. In practice, we construct generators with
$g=2^m$, where $m$ is a positive integer, and generators
with $g=p=2^m-1$, where $m$ is a Mersenne exponent,
i.e., $p=2^m-1$ is a prime.

The notion of an RNG can be formalized as follows: a generator is a
structure ${\cal G}=(S,s_0,T,U,G)$, where $S$ is a finite set of {\it
states}, $s_0\in S$ is the {\it initial state} (or {\it seed}), the
map $T: S\rightarrow S$ is the {\it transition function}, $U$ is a
finite set of {\it output} symbols, and $G:S\rightarrow U$ is the {\it
output function}~\cite{Lecuyer}. Thus, the state of the generator is
initially $s_0$, and the generator changes its state at each step,
calculating $s_n=T(s_{n-1})$, $u_n=G(s_n)$ at step $n$. The values
$u_n$ at the output of the generator are called the {\it observations}
or the {\it random numbers} produced by the generator. The output
function $G$ may use only a small part of the state information
to calculate the random number, the majority of the information being
ignored. In this case, there exist {\it hidden variables}, i.e., some part
of the state information is ``hidden'' and cannot be restored using only
the sequence of RNG observations.

We consider the generator with $S=L^s$, where
$L=\{0,1,\dots,g-1\}\times\{0,1,\dots,g-1\}$ is the lattice on the
torus and $s$ is a positive integer. In other words, the state
consists of coordinates of $s$ points of the $g{\times}g$ lattice
on the torus. For instance, the initial state consists of points
${x_i^{(0)}\choose y_i^{(0)}}$, where $x_i^{(0)}, y_i^{(0)} \in
\{0,1,\dots,g-1\}$ and $i=0,1,\dots,(s-1)$. We note that these are
points of the integer lattice, i.e., $x_i^{(0)}$ and $y_i^{(0)}$ are
positive integers. The actual initial points on the unit
two-dimensional torus $(0,1]\times(0,1]$ are

\begin{equation}
{x_i^{(0)}/g\choose y_i^{(0)}/g},\quad  i=0,1,\dots,(s-1).
\end{equation}

The transition function of the generator is defined by
the action of the cat map $R$, i.e.,
these $s$ points are affected at every step by the cat map:
\begin{equation}
{x_i^{(n)}/g\choose y_i^{(n)}/g}=M
{x_i^{(n-1)}/g\choose y_i^{(n-1)}/g}\mod{1},\quad i=0,1,\dots,(s-1).
\end{equation}
Here the $\bmod\ 1$ operation means taking the fractional part in $(0,1)$
of the real number. An equivalent description of the transition function
is
\begin{equation}
{x_i^{(n)}\choose y_i^{(n)}}=M
{x_i^{(n-1)}\choose y_i^{(n-1)}}\mod{g}, \quad i=0,1,\dots,(s-1).
\label{MatrixRecurr}
\end{equation}

We let $\alpha_i^{(n)}$ denote $0$ or $1$ depending on
whether $x_i^{(n)}<(g/2)$ or $x_i^{(n)}\ge (g/2)$, i.e.,
$\alpha_i^{(n)}=\lfloor 2x_i^{(n)}/g\rfloor$.
The output function of the generator
$G:L^s\rightarrow \{0,1,\dots,2^s-1\}$
is defined as $a^{(n)}=\sum_{i=0}^{s-1} \alpha_i^{(n)}\cdot 2^i$.
In other words, $a^{(n)}$ is an $s$-bit integer consisting of
the bits $\alpha_0^{(n)},\alpha_1^{(n)},\dots,\alpha_{s-1}^{(n)}$.
In the case $g=2^m$, $a^{(n)}$ contains precisely the first
bits of the integers $x_0^{(n)},x_1^{(n)},\dots,x_{s-1}^{(n)}$.
The sequence of random numbers produced by the generator
is $\{a^{(n)}\}$.

We see that the constructed RNG has much hidden information.
For example, if $g=2^m$, then $s(m{-}1)$ bits of
${x_i^{(n)}\choose y_i^{(n)}}$ are the hidden variables;
these are the bits that are not involved in constructing
the value of the output function $a^{(n)}$.

Thus, applying the chaotic behavior of Anosov motion and introducing
an ensemble of systems while keeping part of the information hidden
are the main ingredients of the proposed method.
Good stochastic properties of the underlying continuous system are obviously
necessary for good generators. For example, the logarithm of the multiplier in
the continuous transformation of the LCG can be viewed as the Lyapunov exponent,
which is always greater than $1$, and this leads to the divergence of
trajectories. The huge number of points on a lattice makes the continuous
system a good first approximation to the RNG and leads to the importance of good
chaotic properties. Introducing hidden variables reduces correlations
(as is shown in Sec.~\ref{StatTestsSec}).

The calculation of the period of the RNG is presented in
Appendix~\ref{RNGPeriod}. The typical period length is
$T_m=3\cdot 2^{m-2}$ for the $2^m{\times}2^m$ lattice.
The proper initializations for the generators are
presented in Appendix~\ref{RealizationSec}. The proper
initialization guarantees that the actual period is not smaller
than $T_m$ and that the points ${x_i^{(0)}\choose y_i^{(0)}}$,
$i=0,1,\dots,(s-1),$ belong to different orbits of the cat map.

\subsection{Connection with other generators}

There are several known connections between Anosov dynamical systems and
pseudorandom number generation.

First, the concept of the Shift Register Sequence, which is
widely used to construct high-quality RNGs, is connected to
dynamical systems (see, e.g., the discussion in~\cite{SB-dyn}).
Let the state of the shift register be ${\bf
v_{n-1}}=(a_{n-r},a_{n-r+1},\dots,a_{n-1})$. At the next iteration,
the state of the shift register is ${\bf
v_n}=(a_{n-r+1},a_{n-r+2},\dots,a_n)$, where
$a_n=c_ra_{n-r}+c_sa_{n-s} \mod{2}$. In other words, ${\bf
v_{n+1}}=A{\bf v_n}\mod{2}$, where $A$ is an $(r{\times}r)$-matrix.

Second, LCGs in some cases can be described
by the action of the hyperbolic toral automorphism~\cite{Bonelli}.

Last, it can be shown that, for each $i$, the sequence $\{x_i^{(n)}\}$,
defined above, as well as the sequence $\{y_i^{(n)}\}$, follows a
linear recurrence modulo $g$:

\begin{eqnarray}
\label{Recurr}
x^{(n)}=kx^{(n-1)}-qx^{(n-2)} \mod{g} \\
y^{(n)}=ky^{(n-1)}-qy^{(n-2)} \mod{g},
\end{eqnarray}
where $k=\mbox{Tr} (M)$, and $q=\det M=1$. The characteristic polynomial
of the last linear recurrence is $f(x)=q-kx+x^2$, which is exactly
the same as that of the matrix $M$~\cite{Grothe,Lecuyer90}.

The period properties of sequence (\ref{Recurr}) follow
from the arithmetical methods for $q=1$ (see Appendix~\ref{CatMapsPeriod}
and Appendix~\ref{RNGPeriod}) and from the finite field theory in the
case where $g=p$ is a prime~\cite{Knuth}.

\subsection{Generators for prime $g$: modifications for
$\det M=1$ and for $\det M\ne 1$}

The matrix generator of pseudorandom numbers equivalent
to sequence (\ref{Recurr}) was studied
in~\cite{Knuth,Grothe,NiederrBook} in the case where $g=\det M=p$
is a prime. Sequence (\ref{Recurr}) yields the maximum
possible period $p^2{-}1$ if and only if the characteristic
polynomial $f(x)$ is primitive over $\Z_p$. But for $q=1$
the polynomial $f(x)=x^2-kx+1$ is not primitive over $\Z_p$ for
$p>2$. Therefore, if $\det M=1$, the period is always smaller than
$p^2-1$. An even stronger result follows
from~\cite{PercivalVivaldi}: the period cannot be larger than
$p+1$ when $g$ is a prime  and $\det M=1$.

Matrix generators with $q\ne 1$ are not immediately connected with
Hamiltonian dynamical systems. Indeed, the transformation with $\det M\ne
1$ does not preserve the volume in phase space and does not immediately
represent a cat map. However, whatever $q$ is, we have $\det M^{p-1}\equiv
1\mod{p}$. This means that the action of the matrix $M^{p-1}$ on a lattice
$p\times p$ is exactly the same as the action of a unimodular matrix.
Therefore, any orbit of a ``non-Hamiltonian'' transformation $M$ contains
exactly $p-1$ cat-map orbits.

Also, transformations with $q=1$ preserve the norm on the orbit modulo $g$
(see Appendix~\ref{NormDiscussion}), in contrast to transformations with
$q\ne 1$. It is shown in Appendix~\ref{NormDiscussion} that some of the
correlations between the orbits are inherent in the case $q=1$ and
are suppressed for $q\ne 1$.

\subsection{Rotating the RNG output}
\label{RevolvingSec}

It will be seen that in the scheme of the generator, there are
correlations between the first bits of $a^{(n)}$, correlations
between the second bits of $a^{(n)}$, and so on. To
suppress these correlations, we modify the algorithm as follows.
At each step, we renumber the points in the generator output:
$1\rightarrow 2, 2\rightarrow 3, \dots, s\rightarrow 1$. In other
words, the bits inside $a^{(n)}$ are rotated, and the RNG output
function is defined as
$b^{(n)}=\sum_{i=0}^{s-1}\alpha_i^{(n)}\cdot 2^{(i+n)\mod{s}}$
instead of $a^{(n)}=\sum_{i=0}^{s-1}\alpha_i^{(n)}\cdot 2^{i}$,
where $\alpha_i^{(n)}=\lfloor 2x_i^{(n)}/g\rfloor$.

The main advantage of the modified algorithm is that
it leads to decreasing the correlations of the values
$a^{(n)}$ between each other. For example, we will see in
Sec.~\ref{RandWalksSec} that the rotation strongly reduces
the specific correlations found by the random walks test.

We note that the rotating the bits in the RNG output does not deteriorate
any properties of the RNG provided that $s$ divides the period of
free orbits $T_m$ (in practice, this is a very realistic condition).
In particular, neither does the generator period become smaller
(see Appendix~\ref{RNGPeriod}), nor do the statistical properties
become worse.

Rotating the bits in the RNG output is thus a practically
useful modification. In addition, the rotation makes deciphering
an even more complicated problem.

\section{Statistical tests}
\label{StatTestsSec}
\subsection{Simple Knuth tests}
\label{KnuthStatTests}

In this section, we present the results of several standard
statistical tests~\cite{Knuth} that reveal the correlation properties
of the generator described in Sec.~\ref{GeneratorSec}.
Namely, the frequency test, serial test, maximum-of-t test,
test for monotonic subsequences (``run test'') and collision test
were applied for an RNG with $M={2\ 3\choose3\ 5}$,
$g=2^m=2^{28}$ and $s=28$ points in the state. All the statistical tests
were passed. All empirical tests except the collision test (CT) are based
on either the chi-square test ($\chi^2$) or the Kolmogorov--Smirnov test (KS).
We follow Knuth's notation~\cite{Knuth}.

The results of the tests are presented in Table~\ref{StatTestsTable},
where $n$ is the number of values of $a^{(n)}$ for each test (for the
serial test, $n$ is the number of pairs $\{a^{(2n)},a^{(2n+1)}\}$) and
$\nu$ is the number of degrees of freedom. For the serial test
$d=8$, i.e., we used exactly $3$ bits of each $a^{(n)}$ number;
hence, $\nu=d^2-1=63$. For the run test, $\nu=5$ means that we sought
monotonic subsequences of lengths 1,2,3,4,5 and of length $\ge6$.

\begin{figure}[htb]
\includegraphics[width=0.49\textwidth]{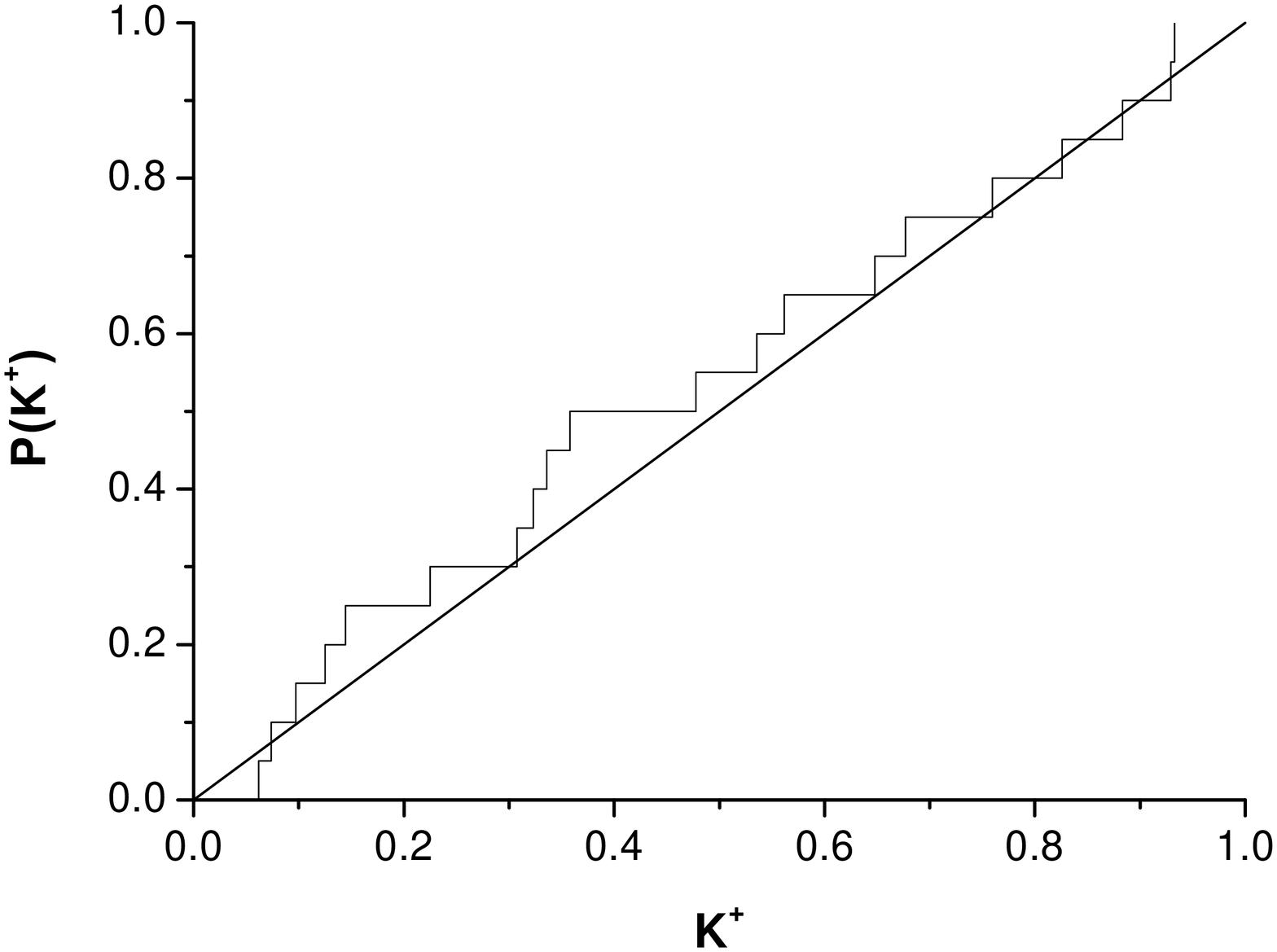}
\includegraphics[width=0.49\textwidth]{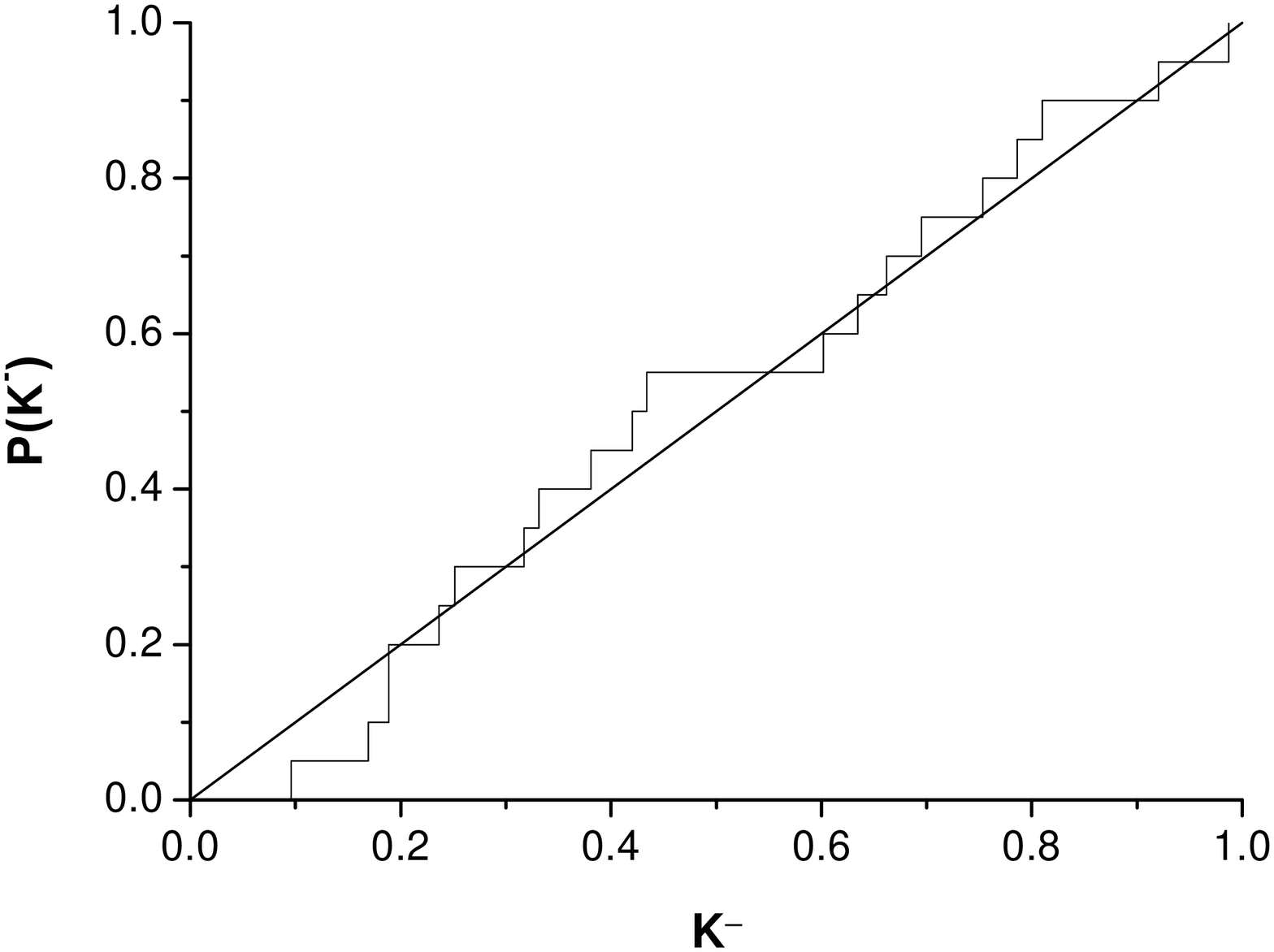}
\caption{Distribution of $P(K^+)$ and $P(K^-)$ for the frequency test.
The test was performed for the RNG from Sec.~\ref{GeneratorSec} with the
parameters $M={2\ 3\choose3\ 5}$, $g=2^m=2^{28}$, $s=28$.}
\label{test}
\end{figure}

For all of the KS tests, the empirical distributions of $P(K^+)$ and
$P(K^-)$ were calculated, where $P(x)$ is the theoretical
Kolmogorov--Smirnov distribution~\cite{Knuth}. Figure~\ref{test} shows
these empirical distributions for the frequency test. These
distributions lead to their own values of $K^+$ and $K^-$: the values
$P(K'^+)$ and $P(K'^-)$ characterizing the empirical distribution of
$K^+$ and the values $P(K'^+)$ and $P(K'^-)$ characterizing the empirical
distribution of $K^-$. These values are presented in
Table~\ref{StatTestsTable}. Our RNG passes all the KS tests because the
values $K^+$ and $K^-$ are distributed in accordance with the theory
prediction. For each chi-square test, the empirical distribution of 20
values of $P(V)$ was calculated. In our tests, it looks similar to
those shown in Fig.~\ref{test}, where $P(x)$ is the theoretical
chi-square distribution and $V$ is the output of the chi-square test. For
each collision test, the number of collisions $c$ and the theoretical
probability $P(c)$ that the number of collisions is not larger than
$c$ were calculated. The empirical distribution of $P(c)$ was analyzed, and
the results are also presented in Table~\ref{StatTestsTable}.

\begin{table}[hbt]
\caption{Results of the statistical tests for the RNG based on the
ensemble of cat maps (see Sec.~\ref{GeneratorSec}) with parameters
$M={2\ 3\choose3\ 5}$, $g=2^m=2^{28}$, $s=28$.}
\begin{tabular}{|l|l|c|c|c|c|c|c|c|c|}
\hline
Test & Parameters & \multicolumn{2}{|c|}{Number and}&
Tests output values  &
\multicolumn{2}{|c|}{Distribution of $V_1$} &
\multicolumn{2}{|c|}{Distribution of $V_2$} &
Conclusion \\
\hhline{|~|~|~|~|~|-|-|-|-|~|}
&&\multicolumn{2}{|c|}{ type of tests} &
$V_1$ and $V_2$ &$P(K'^+)$&$P(K'^-)$&$P(K'^+)$&$P(K'^-)$& \\
\hline
Frequency test & $n=10^6$ &\verb# #$20$\verb# # & KS &
$V_1=K^+, V_2=K^-$&
$0.592392$ & $0.174451$ & $0.457758$ & $0.473379$ & PASSED \\
\hline
Serial test & $n=10^6,d=8$ & $20$ &$\chi^2$ & $V_1=V$ &
$0.837112$ & $0.17128$ &n/a&n/a& PASSED \\
\hline
Run test & $n=10^6, \nu=5$ & $20$ &$\chi^2$ & $V_1=V$ &
$0.383601$ & $0.805434$ &n/a&n/a& PASSED \\
\hline
Maximum-of-t test & $n=10^6,t=5$ & $20$ & KS &  $V_1=K^+, V_2=K^-$&
$0.120912$ & $0.765201$ & $0.704026$ & $0.589702$ & PASSED \\
\hline
Collision test & $m=2^{20}, n=2^{14}$ & $20$ & CT & $V_1=c$&
$0.150537$ & $0.858955$ &n/a&n/a& PASSED \\
\hline
\end{tabular}
\label{StatTestsTable}
\end{table}

We note that all the empirical tests here except the collision test
are essentially multibit. This means that the whole ensemble of cat maps
influences the test result, and one can guess that
hidden variables inside the generator is one reason for the
successful test results.
A single-bit cat-map generator, i.e., a generator from
Sec.~\ref{GeneratorSec} with $s=1$, does not contain hidden
variables.
Most of the tests for a single-bit cat-map generator are also
successfully passed. Namely,
the frequency test and the serial test, which were modified
for a one-bit generator, and the collision test are passed. But there
are correlations in the single-bit cat-map generator (discussed later
in this paper), and the most convenient method for observing them is
the random walks test
with $\mu=1/2$ (see Sec.~\ref{RandWalksSec}). The random walks test
is not the only test that can reveal the single-bit cat-map correlations.
The same correlations are also observed by improved versions of some of the
standard tests, e.g., the serial test for subsequences of length 5.
Of course, many tests in Sec.~\ref{Batteries} would not be passed
by a single-bit cat-map generator.

For comparison, we analyzed a simple generator based on the single cat
map. Table~\ref{SingleCatMapStatTests} shows that such a generator with the
transition function defined as ${x^{(n)}\choose
y^{(n)}}=M{x^{(n-1)}\choose y^{(n-1)}}\mod{1}$ and the output function
defined as $u_n=x^{(n)}$ has very bad properties. Of course, the
frequency test is passed, since the trajectories of a cat map uniformly
fill the phase space. But all the other tests are failed. Therefore,
the simple generator based on the single cat map does have strong
correlations in the output and  is not useful practically.

\begin{table}[hbt]
\caption{Results of the statistical tests for a simple RNG based on the
single cat map with parameters $M={2\ 3\choose 3\ 5}$, $g=2^m=2^{28}$.}
\begin{tabular}{|l|l|c|c|c|c|c|c|c|c|}
\hline
Test & Parameters & \multicolumn{2}{|c|}{Number and}&
Tests output &
\multicolumn{2}{|c|}{Distribution of $V_1$} &
\multicolumn{2}{|c|}{Distribution of $V_2$} &
Conclusion \\
\hhline{|~|~|~|~|~|-|-|-|-|~|}
&&\multicolumn{2}{|c|}{ type of tests} &
values $V_1$, $V_2$ &$P(K'^+)$&$P(K'^-)$&$P(K'^+)$&$P(K'^-)$& \\
\hline
Frequency test & $n=10^7$ &\verb# #$20$\verb# # & KS &
$K^+; K^-$&
$0.96235$ & $0.170581$ & $0.787296$ & $0.067341$ & PASSED \\
\hline
Serial test & $n=5\cdot 10^6, d=8$ & $20$ &$\chi^2$ & $V$ &
$0.989499$ & $0.006013$ &n/a&n/a& FAILED \\
\hline
Run test & $n=10^6, \nu=5$ & $20$ &$\chi^2$ & $V$ &
$0$ & $1$ &n/a&n/a& FAILED \\
\hline
Maximum-of-t test & $n=10^6,t=5$ & $20$ & KS &  $K^+; K^-$&
$0$ & $1$ & $0$ & $1$ & FAILED \\
\hline
Collision test & $d=4,m=2^{10},n=2^{14}$ & $20$ & CT & $c$&
$0$ & $1$ &n/a&n/a&FAILED \\
\hline
\end{tabular}
\label{SingleCatMapStatTests}
\end{table}

\subsection{Batteries of stringent statistical tests}
\label{Batteries}

Knuth tests are very important but still not sufficient for the
present-day sound analysis of the RNG statistical properties. Hundreds
of statistical tests and algorithms are available in software
packages, for example, widely used packages DieHard~\cite{Diehard},
NIST~\cite{NIST} and TestU01~\cite{TestU01}. All of them include
tests, described by Knuth~\cite{Knuth}, as well as many other tests.

Table~\ref{BatteriesOfTests} shows the summary results for the SmallCrush,
PseudoDiehard, Crush and Bigcrush batteries of tests from~\cite{TestU01}.
SmallCrush, PseudoDiehard, Crush and Bigcrush contain 14, 126, 93
and 65 tests respectively. The detailed parameters and initializations
for the generators GS, GR, GSI, GRI, GM19 and GM31, based on the scheme proposed
in Sec.~\ref{GeneratorSec}, are given in Appendix~\ref{RealizationSec}.

For comparison, we also test several other generators, namely, the
standard generators RAND, RAND48 and RANDOM and the modern generators MT19937,
MRG32k3a and LFSR113. RAND is the simple LCG generator based on the recursion
$x_n=(1103515245\; x_{n-1}+12345)\mod{2^{31}}$. RAND48 is the 64-bit LCG
based on the recursion $x_n=25214903917\; x_{n-1}+11\mod{2^{48}}$. RANDOM
provides an interface to a set of five additive feedback random number
generators. RAND, RAND48 and RANDOM are implemented in the functions
\verb#rand()#, \verb#rand48()# and \verb#random()# in the standard Unix or
Linux C library \verb#stdlib# (see the documentation to \verb#rand()#,
\verb#rand48()# and \verb#random()#). MT19937 is the 2002 version of
the Mersenne Twister generator of Matsumoto and Nishimura~\cite{MT}, which is
based on the recent generalizations to the GFSR method.
MRG32k3a is the combined multiply recursive generator
proposed in~\cite{CombinedLCG}, and LFSR113 is a
combined Tausworthe generator of L'Ecuyer~\cite{LFSR113}.

The detailed statistics for the batteries of tests and the explicit results
for every single test from the batteries can be found in~\cite{AlgSite}.

\begin{table}[hbt]
\caption{Numbers of failed tests for the batteries of tests
SmallCrush, Crush, Bigcrush~\cite{TestU01}, and
DieHard~\cite{Diehard}. Here $k=\mbox{Tr} (M)$ and $q=\det M$ are
the RNG parameters (see Sec.~\ref{GeneratorSec} and
Appendix~\ref{RealizationSec}). For each test, we present three
numbers: the number of statistical tests with p-values outside the
interval $[10^{-2},1-10^{-2}]$, number of tests with p-values
outside the interval $[10^{-5},1-10^{-5}]$, and number of tests
with p-values outside the interval $[10^{-10},1-10^{-10}]$.}
\begin{tabular}{|l|c|c||c|c|c|c||}
\hline
Generator & $k$ & $q$ & SmallCrush & Diehard & Crush & Bigcrush \\
\hline
GS       & $3$  & $1$  & $0,0,0$ & $44,29,29$ & $20,16,14$ & $22,20,19$ \\
GR       & $3$  & $1$  & $0,0,0$ & $ 5, 0, 0$ & $ 5, 1, 0$ & $15,10, 7$ \\
GSI      & $11$ & $1$  & $0,0,0$ & $ 1, 0, 0$ & $10, 1, 0$ & $13, 7, 6$ \\
GRI      & $11$ & $1$  & $1,0,0$ & $ 6, 0, 0$ & $ 5, 0, 0$ & $13, 6, 5$ \\
GM19     & $15$ & $28$ & $0,0,0$ & $ 2, 0, 0$ & $ 2, 0, 0$ & $ 3, 0, 0$ \\
GM31     & $7$  & $11$ & $0,0,0$ & $ 2, 0, 0$ & $ 3, 0, 0$ & $ 1, 0, 0$ \\
\hline
RAND     & $-$ & $-$ & $13,13,12$ & $88,84,82$ & $102,100,100$ & $85,83,79$ \\
RAND48   & $-$ & $-$ & $ 5, 5, 3$ & $27,23,22$ & $  22,20,20 $ & $27,23,22$ \\
RANDOM   & $-$ & $-$ & $ 3, 2, 2$ & $17,15,15$ & $  13,11,10 $ & $21,15,14$ \\
MRG32k3a & $-$ & $-$ & $ 1, 0, 0$ & $ 3, 0, 0$ & $   4, 0, 0 $ & $ 2, 0, 0$ \\
LFSR113  & $-$ & $-$ & $ 0, 0, 0$ & $ 3, 0, 0$ & $   8, 6, 6 $ & $ 8, 3, 3$ \\
MT19937  & $-$ & $-$ & $ 0, 0, 0$ & $ 2, 0, 0$ & $   1, 0, 0 $ & $ 4, 0, 0$ \\
\hline
\end{tabular}
\label{BatteriesOfTests}
\end{table}

We consider the test ``failed'' if the p-value lies
outside the region $[10^{-2},1-10^{-2}]$.
Most of the p-values for the failed tests for the cat-map
generators are of the order of $10^{-3}$ to $10^{-5}$,
but several are very small.
We believe that the reason for small p-values is connected with
the small period of the generators GS, GR, GSI, GRI and UNIX RAND.
A period of the order of $3\cdot 10^9$, while sufficient
for some applications, is not sufficient for many of the tests
from Crush and Bigcrush.
Therefore, the generators GS, GR, GSI and GRI demonstrate smaller p-values
and larger numbers of failed tests from Crush and Bigcrush.

The existence of linear congruential dependences between orbits
is another reason for small p-values for GS, GR, GSI and GRI.
These correlations are described analytically in
Appendix~\ref{NormDiscussion}. The GM19 and GM31 generators,
having a period sufficient for the Crush and Bigcrush batteries,
are simultaneously free from the linear congruential dependences.
Therefore, they demonstrate much better statistical properties
in Table~\ref{BatteriesOfTests}.

Because we apply hundreds of tests, the number of failed tests
is susceptible to random statistical flukes, especially when
the p-values of failed tests lie in the suspect region
$[10^{-2},10^{-5}]\cup [1-10^{-2},1-10^{-5}]$.
Table~\ref{TestingTwiceGM31} illustrates the flukes by showing the results
of all batteries of tests for the generator GM31. The batteries were executed
in the order SmallCrush, SmallCrush, PseudoDiehard,
PseudoDiehard, Crush, Crush, BigCrush and BigCrush, i.e. each
battery was executed twice. For the tests in
Tables~\ref{BatteriesOfTests} and~\ref{TestingTwiceGM31}, the generator GM31
was initialized with identical parameters, in accordance with
Appendix~\ref{RealizationSec}.
The numbers of failed tests themselves
in Table~\ref{BatteriesOfTests} and Table~\ref{TestingTwiceGM31}
approximately indicate the statistical robustness of the generators.
But if the p-value lies in the suspect region, one does not know
exactly whether systematic correlations were found in the RNG
or a statistical fluke occured.

\begin{table}
\caption{Applying each battery of tests twice for the GM31 generator.
Here the test is considered failed if the p-value lies
outside the interval $[10^{-2},1-10^{-2}]$}.
\begin{tabular}{|c|c||c|c|c||c|c|c||}
\hline
& Number of & \multicolumn{3}{|c||}{Failed tests, testing first time}
&\multicolumn{3}{|c||}{Failed tests, testing second time}\\
\cline{3-8}
& failed Tests & No & Name & p-value & No & Name & p-value \\
\hline
SmallCrush       & 0/0 &$-$&$-$&$-$&$-$&$-$&$-$\\
\hline
PseudoDiehard    & 3/2  & 1 & BirthdaySpacings & $0.0071$& 6 &CollisionOver& $0.0014$\\
                   &&    6 & CollisionOver & $0.999$&7 &CollisionOver& $0.0018$\\
                   &&    14 & Run of U01 & $0.0093$&&& \\
\hline
Crush            & 2/2 & 70 &Fourier1, $r=0$&  $0.0095$& 14 &BirthdaySpacings, $t=7$& $0.0024$\\
                   &&    71 &Fourier1, $r=20$& $0.0093$& 70 &Fourier1, $r=0$&  $0.0033$\\
\hline
BigCrush         & 4/3 & 6  &MultinomialBitsOver&$0.0014$& 32 &SumCollector&$0.0067$\\
                   &&    23 &Gap, $r=0$&$0.9970$& 36 &RandomWalk1 J(L=90)&$0.9952$\\
                   &&    27 &CollisionPermut&$0.0052$&    41 &RandomWalk1 J(L=10000)&$0.9965$\\
                   &&    39 &RandomWalk1 H(L=1000)&$0.9982$&&&\\
\hline
\end{tabular}
\label{TestingTwiceGM31}
\end{table}

We conclude that the best of the generators based on cat maps are
competitive with other good modern generators. In particular, we recommend
the RNG realizations for GRI and GM31 for practical use.
In Appendix~\ref{RealizationSec}, we present the
effective realizations of the generators GRI-SSE and GM31-SSE and recipes for
the proper initialization. Among the generators examined here these are the best
respective realizations with $g=2^m$ and with prime $g$.

\section{The Random Walks Test}
\label{RandWalksSec}

Analyzing the statistical properties of the generator theoretically
is another important challenge. Such an analysis is traditionally performed
by discussing the lattice structure~\cite{Lecuyer98,Lecuyer90,
AfferbachGrothe} and discussing the discrepancy~\cite{NiederrRev}.
The discrepancy of a matrix generator was analyzed
by Niederreiter~\cite{NeiderrSerial}, who in particular,
proved that the behavior of the discrepancy is
strongly connected to the behavior of an integer called the {\it figure
of merit}.
Although calculating the exact values of the figure of merit
would give an excellent basis for the practical selection of matrices
for matrix generators, these values are still very hard to compute.
To the best of our knowledge, this calculation has never been done
for matrix generators of pseudorandom numbers.

Because of the hidden variables, the lattice structure of the matrix
generator does not directly influence the statistical properties of
the RNG introduced in Sec.~\ref{GeneratorSec}.
Instead, we seek other kinds of correlations
using the random walks test.
The random walks test proved sensitive and powerful for revealing
correlations in RNGs. In particular, correlations in the shift register
RNG were found~\cite{SR2} and explained~\cite{SB,RandomWalksTest} using
the random walks tests. In addition, the random walks test is a
useful tool for analysis: if it fails, it gives the opportunity
to understand the nature of the correlations for a particular RNG~\cite{S99}.

There are several variations of the random walks test in different
dimensions~\cite{MonteCarloBook}. We consider the one-dimensional
directed random walk model~\cite{RandomWalksTest}: a walker starts at
some site of an one-dimensional lattice, and at discrete times $i$, he
either takes a step in a fixed direction with probability $\mu$ or
stops with probability $1-\mu$. In the latter case, a {\it new} walk
begins. The probability of a walk of length $n$ is
$P(n)=\mu^{n-1}(1-\mu)$, and the mean walk length is $\langle
n\rangle=1/(1-\mu)$. We note that the Ising simulations using cluster
updates with the Wolff method are closely related to the random
walk problem~\cite{SB}. Namely, the mean cluster size in the Wolff
method equals the mean walk length for $\mu=\tanh(J/k_B T)$, where
$J$ is the strength of the spin coupling and $T$ is the temperature.

\begin{figure}[htb]
\includegraphics[width=0.5\textwidth]{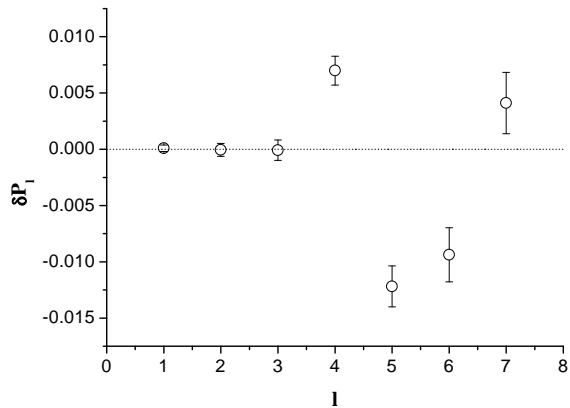}
\caption{The deviation $\delta P_l$ of the probability of a walk length $l$
from the value for uncorrelated random numbers versus walk length $l$.
The mean and the variance for $\delta P_l$ are represented
for 100 chi-square random walk simulations. See the text for the details.}
\label{walktest}
\end{figure}

Figure~\ref{walktest} shows the correlations in the RNG found by the
random walks test. We applied 100 chi-square tests. Each test
performed $n=10^7$ random walks with $\mu=1/2$ for a
generator with $M={2\ 3\choose3\ 5}$, $m=32$ and $s=1$. The result of
the test with $\mu=1/2$ is independent of $s$ because only the
first bit of the RNG is taken into account. For each test, the value
$\delta P_l=(Y_l-np_l)/(np_l)$ was calculated for all walk lengths
$l\le 7$. Here $p_l$ is the theoretical probability of the walk length $l$
for uncorrelated random numbers, and $Y_l$ is the simulated number of walks
with length $l$. We note that correlations can be found only for a large
number of random walks (see Table~\ref{RandomWalksTable}), and no
correlations are found even for $n=6\cdot 10^4$ random walks.

These correlations can be explained as follows.
There are 32 five-bit sequences, and they do not have the same frequency
of appearing in the RNG output. We consider one of them, for example, 10011.
Let $X=(0,\frac12]\times(0,1]$ and $Y=(\frac12,1]\times(0,1]$, i.e.,
$X$ and $Y$ are the left and the right halves of the torus.
Let $x$ be the initial point $x_0^{(0)} \choose y_0^{(0)}$
of the generator.
For the first bits of the first five outputs of the generator
to be 10011, it is necessary and sufficient to have
$x\in Z_{10011}=Y \cap R^{-1}(X) \cap R^{-2}(X) \cap R^{-3}(Y) \cap R^{-4}(Y)$.
Here, $R$ is the action of the cat map.
The set $Z_{10011}$ consists of filled polygons. Each polygon can be
calculated exactly. The area $S(Z_{10011})$ equals the probability for the
first five outputs of the generator to be 10011.
This shows that the nature of the correlations is found in the geometric
properties of the cat map.

\begin{figure}[hbt]
\includegraphics[width=0.46\textwidth]{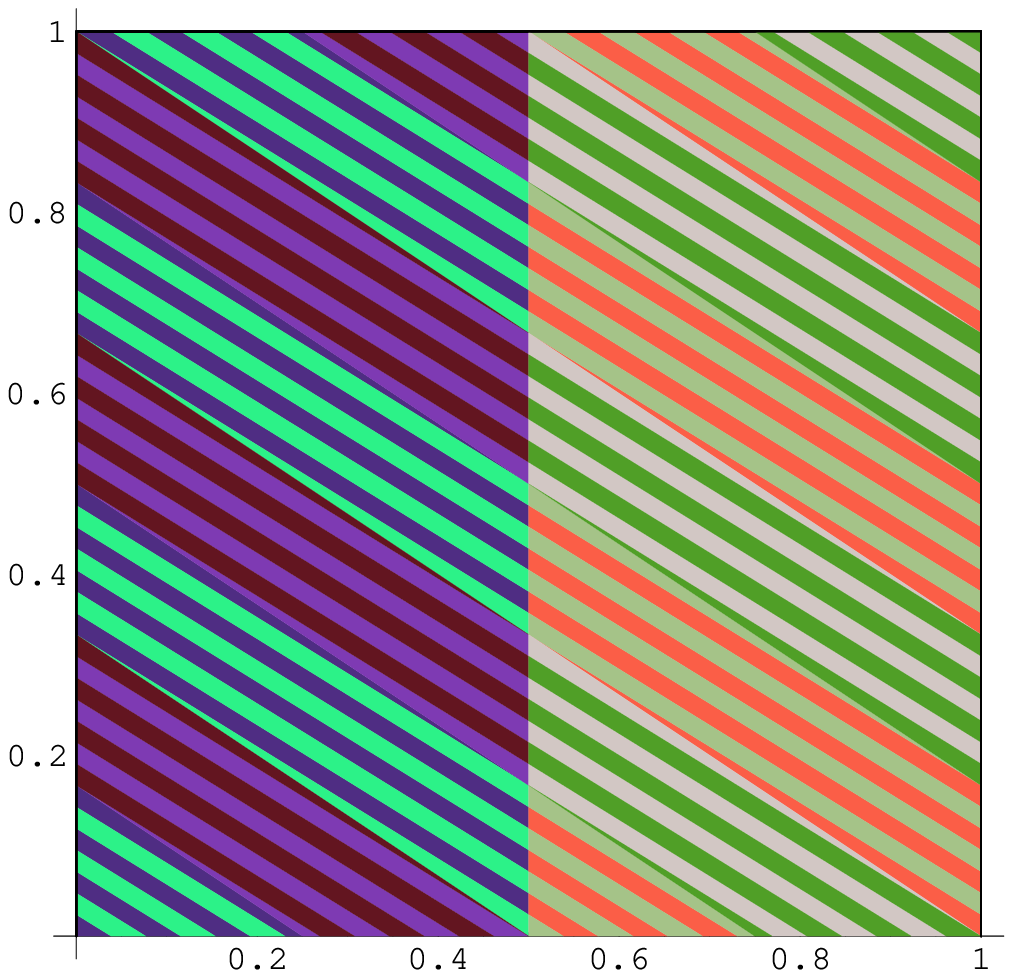}
\includegraphics[width=0.46\textwidth]{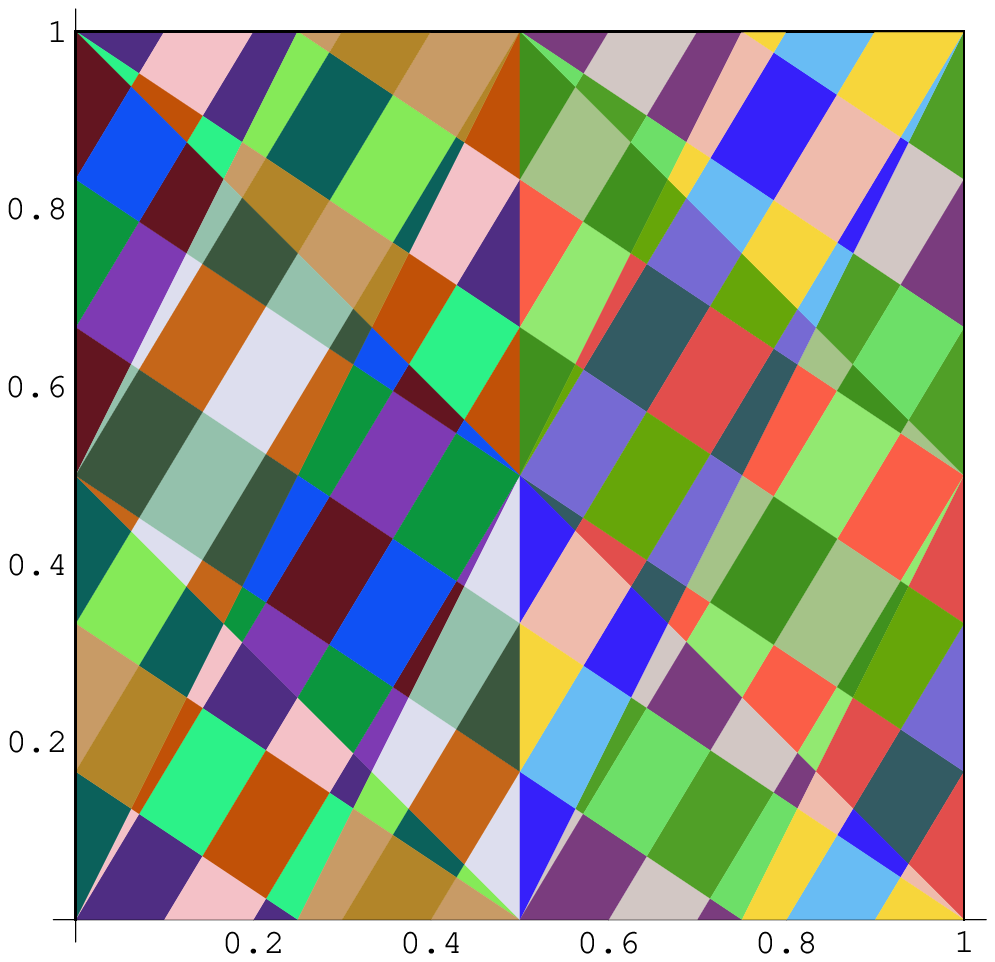}
\caption{(Color online) Left: the regions on the torus for the first
initial point of the RNG described in Sec.~\ref{GeneratorSec} with $M={2\
3\choose3\ 5}$. These regions correspond to the sequences
$000,001,010,011,100,101,110$ and $111$ of the first bits generated by the RNG.
Each region is drawn with its own color. Right: the regions on the
torus for the third point of the RNG described in
Sec.~\ref{GeneratorSec} with $M={1\ 1\choose1\ 2}$. These regions
correspond to the sequences of length 5 of the first bits generated by the
RNG. Each region is drawn with its own color.}
\label{out3}
\end{figure}

Figure~\ref{out3} (the left picture) represents the polygons
corresponding to the subsequences of length three for the cat map with
$M={2\ 3\choose3\ 5}$. Each set of polygons, e.g., $Z_{010}=X \cap
R^{-1}(Y) \cap R^{-2}(X)$, represents the region on the torus for the
first initial point of the RNG and is drawn with its own color. The
right picture represents the subsequences of length five for the cat map
with $M={1\ 1\choose1\ 2}$. Here, each set of polygons represents the
regions on the torus for the third point of the generator, e.g.,
$\tilde Z_{01001}=R^{-2}(X)\cap R^{-1}(Y)\cap X\cap R(X)\cap R^2(Y)$,
and is drawn with its own color. Of course, $S(\tilde
Z_{01001})=S(Z_{01001})=P(01001)$ because the cat maps are area
preserving. Therefore, the choice of pictures of $Z_i$ or pictures of
$\tilde Z_i$ is unimportant if we only want to calculate the areas.
Thus, the geometric structures in Fig.~\ref{out3} show the regions
of $Z_{000},\dots,Z_{111}$ (the left picture) or $\tilde
Z_{00000},\dots,\tilde Z_{11111}$ (the right picture) and illustrate
the geometric approach to calculating the probabilities.

The exact areas $S(Z_{00000}),\dots,S(Z_{11111})$ can be easily calculated
for various toral automorphisms. We prove the following geometric
propositions:
\begin{enumerate}
\item In any case, every subsequence of length $3$, $2$ or $1$
respectively has the same probability $1/8$, $1/4$ or $1/2$.
\item If $k={\rm Tr}(M)$ is an odd number, then every subsequence
of length $4$ has the same probability $P_0=1/16$.
\item If $k$ is even, then the probability of the subsequence $0000$
depends only on the trace $k$ of matrix $M$ of the cat map. It
equals $P=P_0\cdot k^2/(k^2-1)$, where $P_0=1/16$.
\end{enumerate}
The line of reasoning is presented in~\cite{SpaceProofCite}.
Of course, the probability of the subsequence $0000$ automatically gives the
probabilities of all other subsequences of length 4.
We note that if $k$ is odd, then ideal $\langle 2\rangle$ is inert
(see Appendix~{\ref{OrbitPeriods}}), and the inert case is the easiest
for exact analysis of the RNG period
(see Appendices~\ref{CatMapsPeriod} and~\ref{RNGPeriod}).
The probabilities of the subsequences of length 5
for maps with odd traces and of the subsequences of length 4
for maps with even traces are calculated exactly and shown
in Table~\ref{deviations}.
It can be conjectured from Table~\ref{deviations} that if $k$ is odd,
then the probability of the subsequence $00000$ of length 5 equals
$P_0\cdot (1+1/(3k^2-6))$, where $P_0=1/32$.

\begin{table}[hbt]
\caption{The probabilities of subsequences for different cat maps,
characterized by the trace $k$.}
\begin{tabular}{|r|c||r|c||r|c|}
\hline
$k$& $P(0000)/P_0$ &$k$& $P(0000)/P_0$& $k$& $P(00000)/P_0$ \\
\hline
4 & 16/15           &  30 & 900/899    &    3 & 22/21 \\
6 & 36/35           &  32 & 1024/1023  &    5 & 70/69 \\
8 & 64/63           &  34 & 1156/1155  &    7 & 142/141 \\
10 & 100/99         &  36 & 1296/1295  &    9 & 238/237 \\
12 & 144/143        &  38 & 1444/1443  &   11 & 358/357 \\
14 & 196/195        &  40 & 1600/1599  &   13 & 502/501 \\
16 & 256/255        &  42 & 1764/1763  &   15 & 670/669 \\
18 & 324/323        &  44 & 1936/1935  &   17 & 862/861 \\
20 & 400/399        &  46 & 2116/2115  &   19 & 1078/1077 \\
22 & 484/483        &  48 & 2304/2303  &   21 & 1318/1317 \\
24 & 576/575        &  50 & 2500/2499  &   23 & 1582/1581 \\
26 & 676/675        &  52 & 2704/2703  &   25 & 1870/1869 \\
28 & 784/783        &  54 & 2916/2915  &   27 & 2182/2181 \\
\hline
\end{tabular}
\label{deviations}
\end{table}

The probabilities can thus be approximated as $P/P_0=1+Bk^{-2}$
for large $k$, where $P_0=2^{-n}$ for subsequences of length $n=4,5$.
Here $B=1$ when $k$ is even and $n=4$; $B=1/3$ when $k$ is odd and $n=5$.
We conclude that the deviations found by our implementation of
the random walks test will vanish as the trace $k$ increases.

Table~\ref{RandomWalksTable} shows that using rotation in the RNG output
(see Sec.~\ref{RevolvingSec}) results in suppressing correlations found
by the random walks test. This is not surprising, because
even the one-bit random walks test with $\mu=1/2$ deals with the
ensemble of cat maps when the rotation is used.

\begin{table}
\caption{Left: results of the random walks test (i.e. $100$ chi-square random
walks simulations) for different $n$ and for $\mu=1/2$, $m=32$ and $s=1$.
Right: results of the random walks test for different
$\mu$ and $s$ and for $n=10^6$ and $m=32$. Here
$s\ne 1$ means using rotation in the RNG output
(see Sec.~\ref{RevolvingSec}). Actually, the same one-bit
random walks test is used because $\mu=1/2$.}
\begin{tabular}{|c|c|c|c|}
\hline
$n$  & $P(K'^+)$ & $P(K'^-)$ & Result \\
\hline
$10^4$    & $0.746106$ & $0.594428$ & PASSED \\
\hline
$3\cdot 10^4$  & $0.341899$ & $0.675728$ & PASSED \\
\hline
$6\cdot 10^4$  & $0.307433$ & $0.694282$ & PASSED \\
\hline
$10^5$  & $0.018332$ & $0.966717$ & UNCERTAIN \\
\hline
$3\cdot 10^5$ & $0.0001594$ & $1$ & FAILED \\
\hline
$6\cdot 10^5$ & $0.0001378$ & $1$ & FAILED \\
\hline
$10^6$ & $0$ & $1$ & FAILED \\
\hline
\end{tabular}
\begin{tabular}{|c|c|c|c|c|}
\hline
$\mu$ & $s$ & $P(K'^+)$ & $P(K'^-)$ & Result \\
\hline
$1/4$ & $1$ & $0.235776$ & $0.882413$ & PASSED\\
\hline
$3/4$ & $1$ & $0.299174$ & $0.613382$ & PASSED\\
\hline
$1/8$ & $1$ & $0.087016$ & $0.770549$ & PASSED \\
\hline
$1/16$ & $1$ & $0.67967$ & $0.920998$ & PASSED \\
\hline
\hline
$1/2$ & $2$ & $0.605407$ & $0.344068$ & PASSED \\
\hline
$1/2$ & $3$ & $0.527088$ & $0.645272$ & PASSED \\
\hline
$1/2$ & $4$ & $0.558105$& $0.360828$ & PASSED\\
\hline
\end{tabular}
\label{RandomWalksTable}
\end{table}

\section{Discussion}

In this paper, we have proposed a scheme for constructing a good RNG.
The distinctive features of this approach are
applying the ensemble of cat maps while taking only
a single bit from the point of each cat map and applying
methods that allow analyzing both the properties of the
periodic orbits and the statistical properties of such
a generator both theoretically and empirically. We have seen that the
algorithm in Sec.~\ref{GeneratorSec} can generate sequences with
very large period lengths.
Although essential correlations are always present and
important statistical deficiencies are found, a good algorithm
with proper initialization can minimize them.
The best generators created by this method have statistical properties
that are not worse, and speed is slightly slower than that of good
modern RNGs. The techniques used allow calculating the period
lengths and correlation properties for a wide class of sequences
based on cat maps.

Future modifications and enhancements are possible, and we currently
recommend the generators GM19-SSE, GM31-SSE and GRI-SSE for practical
use. Program codes for the generators and for the proper
initialization can be found in~\cite{AlgSite} and the generator details
are discussed in
Appendix~\ref{RealizationSec}. We would appreciate any comments on
user experiences.

\section{Acknowledgments}

We are grateful to the anonymous referee for the critique and
questions that allowed essentially improving the content of the
paper. This work was supported by the US DOE Office of Science under
contract No. W31-109-ENG-38 and by the Russian Foundation for Basic
Research.

\appendix
\section{Periodic orbits of the cat maps on the $2^n{\times}2^n$ lattice}

In this section, we review the key arithmetic methods for studying
orbit periods that are described in detail in~\cite{PercivalVivaldi}.
Some of the results are presented in this appendix in a more general
form. The notation is discussed briefly; the details and proofs on the
formalism of quadratic integers and quadratic ideals can be found
in~\cite{Cohn, Chapman}.
\label{CatMapsPeriod}

\subsection{The dynamics of the cat map and rings of quadratic integers}

We consider the unit two-dimensional torus
(the square $(0,1]\times(0,1]$ with the opposite sides identified).
We take a cat map
$M=\left(\begin{array}{cc} m_{11}&m_{12}\\m_{21}&m_{22}\\ \end{array}\right)
\in SL_2(\Z)$, which acts on a lattice $g\times g$ on the torus,
where $g=2^n$. The elements of $M$ are integers, $\det M=1$
and $|k|>2$, where $k={\rm Tr} (M)$.

For any given trace $k>2$, there exists a unique map $M\in SL_2(\Z)$
such that the connection between the properties of periodic orbits
of the automorphism and the arithmetic of quadratic integers is
the most natural. Indeed, we consider a matrix $M$ such that
\begin{equation}
\left\{
\begin{array}{rcl}
\lambda &=& m_{11} + \tau m_{21}, \\
\lambda\tau &=& m_{12} + \tau m_{22}. \\
\end{array}
\right.
\label{ChoiceM}
\end{equation}
Here $\tau$ is the base element of the ring of quadratic integers
$R_D=\{a+b\tau: a,b\in\Z\}$ that contains $\lambda$.
This means that $\exists n\in\Z: k^2-4=n^2 D$, where $D$
is a squarefree integer and $\tau=\sqrt{D}$ for $D\not\equiv 1\mod{4}$;
$\tau=\frac12(1+\sqrt{D})$ for $D\equiv 1\mod{4}$.

It easily follows from (\ref{ChoiceM}) that $x'+y'\tau=\lambda(x+y\tau)$
is equivalent to ${x'\choose y'}=M {x\choose y}$ for any $x,y,x',y'$.
Indeed, $\lambda(x+y\tau)=\lambda x+(\lambda\tau)y=
(m_{11}x+m_{12}y)+(m_{21}x+m_{22}y)\tau=x'+y'\tau$.
The action of the map $M$ corresponds to multiplication by the
quadratic integer $\lambda$, while the action of $M^{-1}$
corresponds to multiplication by $\lambda^{-1}$.
Hence, we can choose either of the two eigenvalues
$\lambda=(k\pm\sqrt{k^2-4})/2$, e.g., the largest one,
because the exact choice is unimportant for studying orbit periods.

Generally speaking, there are infinitely many maps in $SL_2(\Z)$ that have
identical eigenvalues, and not all the maps are related by a canonical
transformation (share the same dynamics). But arguments presented
in~\cite{PercivalVivaldi} strongly suggest that they still share the
same orbit statistics.

\subsection{Invariant sublattices on the torus and factoring
quadratic ideals}

We note that each element of $R_D$ represents some point of $\Z^2$.
Let $A$ be a quadratic ideal. We say that $\xi\equiv\eta\mod{A}$
if $(\xi-\eta)\in A$.
We consider the principal quadratic ideal generated by $g$:
$\la g\ra=\{ag+bg\tau: a,b\in\Z\}$. It corresponds to the
set of points of a square lattice with the side $g$.
Then the period of an orbit containing the point
${x/g \choose y/g}$ is the smallest integer $T$
such that $\lambda^T z\equiv z \mod{\la g\ra}$.
Here $x$ and $y$ are integers, and $z=x+y\tau$.

Each quadratic ideal $A$ is associated with some sublattice of $\Z^2$.
Because $\lambda$ is a unit, the sublattice is invariant
with respect to multiplication by $\lambda$: $\lambda A=A$.
Since we are interested in invariant lattices on the unit
two-dimensional torus, we consider only those sublattices of $\Z^2$
that are invariant under an arbitrary translation ${ag\choose bg}$,
where $a,b\in\Z$.
These sublattices correspond to quadratic ideals that divide
$\la g\ra$. Factoring the ideal $\la g\ra$
thus yields invariant sublattices on the torus.

\subsection{The classification of prime ideals and the orbit periods
for the $2^n{\times}2^n$ lattice}
\label{OrbitPeriods}

We consider $2^n{\times}2^n$ lattices on the torus.
Because $\la g\ra=\la 2^n\ra=\la 2\ra^n$, it is sufficient to have the ideal
factorization of $\la 2\ra$. We recall that the ideal $\la 2\ra$ is said
to be {\it inert} if $\la 2\ra$ is already a prime ideal; it is
said to be {\it split} if $\la 2\ra=P_1 P_2$, where $P_1$ and $P_2$ are
prime ideals; it is said to be {\it ramified} if $\la 2\ra=P_1^2$,
where $P_1$ is a prime ideal. The ideal $\la 2\ra$ is inert
for $D\equiv 5\mod{8}$, split for $D\equiv 1\mod{8}$ and ramified for
$D\not\equiv 1\mod{4}$.

It follows that if the trace $k$ is odd, then $\la 2\ra$ is inert;
if $k\equiv 0\mod{4}$, then $\la 2\ra$ is ramified. Indeed, for odd $k$,
we have $k^2-4\equiv 5\mod{8} \Rightarrow D\equiv 5\mod{8}$;
for $k\equiv 0\mod{4}$, we have $(k^2-4)/4=n_1^2 D\equiv 3\mod{4}
\Rightarrow D\equiv 3\mod{4}$. For $k\equiv 2\mod{4}$, we obtain
$(k^2-4)/4=n_1^2 D\equiv 0\mod{4}$, i.e., all three
possibilities (inert, split or ramified ideal $\la 2\ra$) can occur.

Let $T_n$ denote the period of any of the free orbits for $g=2^n$
and $T'_n$ denote the period of those ideal orbits for $g=2^n$
that do not belong to the sublattice $\frac{g}2{\times}\frac{g}2$.
We recall that an orbit belonging to a given lattice $\Z^2/g\Z^2$
is called an {\it ideal orbit} if it belongs to some ideal $A$ such that
$A|\la g\ra$ and $A\ne \la 1\ra$. Otherwise, it is called a {\it free orbit}.

The behavior of periodic orbits on the $2{\times}2$ lattice follows
from Propositions B1--B3 in~\cite{PercivalVivaldi}. Namely, we have
the following:
\begin{itemize}
\item If $\la 2\ra$ is inert, then either $T_1=3$ or $T_1=1$; all orbits
are free.
\item If $\la 2\ra$ is split, then $T_1=T'_1=1$; there are two ideal
orbits and one free orbit.
\item If $\la 2\ra$ is ramified, then $T_1=2$ and $T'_1=1$;
there is an ideal orbit and a free orbit (it is also possible that
$T_1=1$ and $T'_1=1$; there are two free orbits and an ideal orbit).
\end{itemize}

To determine the structure of periodic orbits on
the $2^n{\times}2^n$ lattice, we prove the following theorem.

{\bf Theorem.}
\begin{enumerate}
\item For all $n$, either $T_{n+1}=2T_n$ or $T_{n+1}=T_n$.
\item For all $n$, either $T_n'=T_n$ or $T_n'=T_{n-1}$.
\item For all $n\ge3$, $T_n\ne T_{n-1} \Rightarrow T_{n+1}\ne T_n$.
\item If $n\ge 4$, $T_n\ne T_{n-1}$, and
$T_n'=T_n/a$, where $a \in \{1,2\}$, then $T_{n+1}'= T_{n+1}/a$.
\end{enumerate}

This theorem generalizes Propositions C1 and C2 in~\cite{PercivalVivaldi}.
The line of reasoning is presented in Appendix~\ref{ThProof}.

Therefore, knowing $T_n$ and $T'_n$ for small $n$ suffices
for determining the orbit statistics for all $n$.
There always exist $n_1$, $n_2$ and $n_3$ such that $T_n=T_12^{n-n_1}$
and $T'_n=T_12^{n-n_2}$ for all $n\ge n_3$.

If $\la 2\ra$ is inert, then every ideal that divides $\la g\ra$
has the form $\la 2\ra^r$. Therefore, each ideal orbit belongs
to the $2^{n-1}{\times}2^{n-1}$ sublattice and coincides with
a free orbit for some sublattice $2^r{\times}2^r$, where $r<n$.
We now find the number of free orbits in the inert case.
There are $2^{2n}{-}1$ points on a lattice. The ideal orbits
contain $2^{2n-2}{-}1$ points. Consequently, there are
$(2^{2n}-2^{2n-2})/T_n = 3\cdot 2^{2n-2}/T_n$ free orbits.

We suppose that the typical inert case occurs, i.e., $T_n=3\cdot 2^{n-2}$.
Then the phase space is divided into the following regions:
\begin{itemize}
\item $3/4$ of the phase space is swept by $2^n$ trajectories
of period $T_n$,
\item $3/16$ of the phase space is swept by $2^{n-1}$ trajectories
of period $T_{n-1}=T_n/2$,
\item $3/64$ of the phase space is swept by $2^{n-2}$ trajectories
of period $T_{n-2}=T_n/4$,
\item and so on.
\end{itemize}

All such statements hold as long as
the trajectory length exceeds just a few points.
Therefore, on one hand, cat map orbits have
huge periods; on the other hand, the number of orbits
is sufficiently large (see Theorem 2 in Appendix~\ref{RNGPeriod}). Both
these properties are important for our construction of the RNG.

\section{The RNG period}
\label{RNGPeriod}

In this section, we find the periods of the generators in
Sec.~\ref{GeneratorSec} and Sec.~\ref{RevolvingSec}. As a result
of Appendix~\ref{CatMapsPeriod} and Appendix~\ref{RNGPeriod},
the RNG period can be obtained for arbitrary parameters of the map
and lattice.

{\bf Theorem 1.}
If $g=2^m$, then the period $T$ of the sequence $\{a^{(n)}\}$ in
Sec.~\ref{GeneratorSec} equals the period $T_m$ of free orbits of the cat map
for the overwhelming majority of RNG initial conditions.

{\bf Proof.\ \ }
\begin{enumerate}
\item At least one of the initial
points ${x_i^{(0)}\choose y_i^{(0)}}$ belongs to a free orbit.
Indeed, the probability of this in the inert case equals $(1-4^{-s})$.
\item Therefore, $T$ is not less than $T_m$. Indeed,
$T$ is not less than the period of the sequence of first bits
of $x_i^{(0)}, x_i^{(1)},\dots$ for each $i$.
But the period of the sequence of first bits of
points of the cat map orbit is equal to the orbit period for the vast
majority of orbits. The probability of the opposite
is tiny provided that the orbit is not too short.
\item Finally, $T$ is not larger than $T_m$. Indeed,
the period of each cat map orbit divides $T_m$.
\end{enumerate}

{\bf Example.} In the typical example of the inert case, where
$M={4\ 9\choose3\ 7}$, we obtain $T_m=3\cdot 2^{m-2}$.
This fact was also tested numerically as follows. First, the initial conditions
were set randomly. Second, the period of $\{a^{(n)}\}$ was accurately
found numerically. This operation was repeated 1000 times for $m=s=14$.
Each time the period of $\{a^{(i)}\}$ turned out to be $6144=3\cdot 2^{11}$.
To check the period numerically, we first check whether
the whole state of the RNG (not only the output) coincides at the
moments $0$ and $T$ and then verify that a smaller
period (which could possibly divide $T$) does not exist.

{\bf Theorem 2.}
The probability that two arbitrary points of the $2^m{\times}2^m$ lattice
on the torus belong to the same orbit of the cat map equals
$9/(7\cdot 2^{m+2})$. The probability that $s$ arbitrary points
of the lattice do not belong to $s$ different orbits of the cat map
(i.e., two of the points belong to the same orbit) is
$9s(s{-}1)/(7\cdot 2^{m+3})$.

The proof of Theorem 2 is straightforward. Of course, both these
probabilities are tiny if $m$ is sufficiently large.

{\bf Theorem 3.} If $g=2^m$ and $s|T_m$, then the period $T$ of the
sequence $\{b^{(n)}\}$ in Sec.~\ref{RevolvingSec} equals $T_m$ for the
overwhelming majority of RNG initial conditions.

{\bf Proof.\ \ }
\begin{enumerate}
\item Because $s|T_m$, we have $b_{i+T_m}=b_i$ for all $i$.
Therefore, $T|T_m$.
\item If $s$ does not divide $T$, then
$\forall i\in\{0,1,\dots,s-1\} \exists j\in\{0,1,\dots,s-1\}, j\ne i$,
such that $x_i^{(0)} \choose y_i^{(0)}$ and $x_j^{(0)} \choose y_j^{(0)}$
belong to the same orbit of the cat map. It follows from
Theorem 2 that this event is highly improbable. Therefore, $s|T$.
\item Because $T$ is a period of $\{b^{(n)}\}$ and $s|T$, we have
$b_{i+T}=b_i \Rightarrow a_{i+T}=a_i$ for all $i$. Therefore, $T_m|T$.
\end{enumerate}

The above theorems show that the period calculations for the sequences
$\{a^{(n)}\}$ and $\{b^{(n)}\}$ are reliable in the general case, because
the chance of the period dependence on the initial state is exponentially
small. But it is a desirable property that the period does not depend
on any conditions at all. The proper initialization (see
Appendix~\ref{RealizationSec}) guarantees that (i) at least
one of the initial points belongs to a free orbit of the cat map;
and (ii) no pair of initial points belongs to the same orbit of the cat map.
Therefore, both the periods of $\{a^{(n)}\}$ and of $\{b^{(n)}\}$
are guaranteed to equal $T_m$ provided the initialization
in Appendix~\ref{RealizationSec} is applied.

The above theorems and considerations hold for $g=2^m$. In the other case,
when $g=p$ is a prime, it follows from finite field theory
that the period of any orbit of the matrix transformation is equal
to $p^2-1$ provided the polynomial $f(x)=x^2-kx+q$ is primitive
modulo $p$. The methods for good parameter and initialization choice
for such generators are also presented in Appendix~\ref{RealizationSec} for
the generators GM19 and GM31.
A similar argument as in Theorems 1 and 3 shows that in this case
(i) the period of the sequence $\{a^{(n)}\}$ equals $p^2{-}1$;
and (ii) the period of the sequence $\{b^{(n)}\}$ is divisible by $p^2{-}1$,
i.e., rotation cannot decrease the period of such a generator.

\section{Orbits, Norm and Correlations between orbits}
\label{NormDiscussion}

In this section,
(i) we show that the norm modulo $g$ is the characteristic of the whole orbit;
(ii) we find the number of orbits of each norm modulo $g$ and discuss
how symmetries affect the norm;
and (iii) we find the linear congruential dependences between orbits.
The consideration holds for maps with $q=1$ on a $2^m{\times}2^m$ lattice.

\subsection{Orbits and norm}

We recall that the norm of a quadratic integer $\alpha=a+b\sqrt{D}$ is
simply an integer $N(\alpha)=\alpha\alpha^*=a^2-b^2D$. If $\la 2\ra$ is
inert, then the quadratic integer $x+y\tau$, where $\tau=\frac{1+\sqrt{D}}{2}$,
represents the point $x\choose y$, i.e.,
$N{x\choose y}=x^2+xy-\frac{D-1}{4}y^2$.
A cat map preserves the value of $N{x\choose y}\mod{2^m}$, because the action
of a cat map can be described as $x'+y'\tau=\lambda(x+y\tau)\mod{\la 2^m\ra}$,
where $\lambda$ is a matrix eigenvalue and $N(\lambda)=1$.
Therefore, the norm modulo $g$ is a characteristic of the whole orbit.
We note that for a point on a free orbit, either $x$ or $y$ is odd,
consequently the norm is also an odd number.

We prove that if the period of free orbits is $T=3\cdot 2^{m-2}$,
then for each $N=1,3,\dots,2^m{-}1$,  there are exactly two orbits that have the
norm $N$ (these two orbits are symmetrical, i.e. the second one
contains the points $2^m-x_n\choose 2^m-y_n$, where $x_n\choose y_n$
are the points of the first one).
Indeed, there are exactly $2^m$ free orbits that occupy
$T\cdot 2^m=2^{2m}-2^{2m-2}$ points. On the other hand, there is
a method for obtaining two symmetrical orbits having any odd norm.
We note that other possible symmetries (e.g., symmetries
considered in~\cite{SpaceProofCite}) preserve the norm modulo $256$.
Moreover, in most cases, they preserve the norm modulo $2^{m-1}$ or
modulo $2^{m-2}$.

\subsection{Correlations between orbits}

We consider a pair of free orbits with the norms $N_1$ and $N_2$.
The set $A=\{1,3,\dots,2^m{-}1\}$ is a group under multiplication
(it is called the modulo multiplication group); hence, there exists
$t\in A$ such that $N_1\equiv tN_2\mod{2^m}$. It is known
that for the equation $k^2\equiv t\mod{2^m}$ to have a solution
$k\in A$, it is necessary and sufficient to have $t\equiv 1\mod{8}$.
Thus, if $N_1\equiv N_2\mod{8}$, there exists $k$ such that
$N_1\equiv k^2N_2\mod{2^m}$. If $x_n\choose y_n$ are the points
of the orbit of norm $N_2$, then ${kx_n\choose ky_n}\mod{\la 2^m\ra}$
are the points of the orbit of norm $N_1\equiv k^2N_2\mod{2^m}$, in the
same order. But there may be large shift between the values
of different orbits.

Thus, the case $N_1\equiv N_2\mod{8}$ is dangerous, because
there may be correlations between orbits. The points of the first orbit
are connected to the points of the second orbit with a linear
congruential dependence. The parameter $k$ may be interpreted as
a random odd number.

\section{Proof of the theorem in Appendix~\ref{OrbitPeriods}.}
\label{ThProof}

{\bf Proposition 1.}
$T_n$ is the least integer such that
$\lambda^{T_n}\equiv 1\mod{\la 2^n\ra}$. In particular,
any free orbit has the same period.

{\bf Proof.\ \ }
We suppose that the period of an orbit containing the point $z$
is $T$. Then
$\lambda^Tz\equiv z\mod{\la 2^n\ra} \Rightarrow z(\lambda^T-1)\in \la 2^n\ra$.
If the orbit is free, then $z\not\in P$ for any ideal $P$ such that
$P|\la 2^n\ra$, $P\ne \la 1\ra$.
Therefore, $(\lambda^T-1)\in \la 2^n\ra$.

{\bf Proposition 2.} $T_n|T_{n+1}$.

{\bf Proof.\ \ } Indeed,
$
\lambda^{T_{n+1}}\equiv 1\mod{\la 2^{n+1}\ra} \Rightarrow
\lambda^{T_{n+1}}\equiv 1\mod{\la 2^{n}\ra} \Rightarrow
T_{n+1}=mT_n
$, where $m\in\N$.

{\bf Proposition 3.} For all $n$, either $T_{n+1}=2T_n$ or $T_{n+1}=T_n$.

{\bf Proof.\ \ } Because
$(\lambda^{T_n}-1) \in \la 2^n\ra$, we have
$(\lambda^{T_n}+1)=(\lambda^{T_n}-1)+2 \in \la 2\ra$.
Consequently,
$\lambda^{2T_n}-1=(\lambda^{T_n}-1)(\lambda^{T_n}+1) \in \la 2^{n+1}\ra$,
i.e., either $T_{n+1}=2T_n$ or $T_{n+1}=T_n$.

{\bf Proposition 4.}
If $n\ge 3$ and $T_n\ne T_{n-1}$, then $T_{n+1}\ne T_n$.

{\bf Proof.\ \ } It follows from $T_n=2T_{n-1}$ that
$\left\{
\begin{array}{l}
\lambda^{T_{n-1}} \equiv 1\mod{\la 2^{n-1}\ra} \\
\lambda^{T_{n-1}} \not\equiv 1\mod{\la 2^n\ra} \\
\end{array}
\right.\Rightarrow \lambda^{T_{n-1}}=1+z\cdot2^{n-1}$,
where $z\not\in\la 2\ra$.
Squaring the last equation, we obtain
$
\lambda^{2T_{n-1}}=1+z\cdot 2^n+ z^2\cdot 2^{2n-2}
\equiv 1+z\cdot 2^n \mod{\la 2^{n+1}\ra}$ for $n\ge 3$.
Hence,
$\lambda^{2T_{n-1}}\not\equiv 1\mod{\la 2^{n+1}\ra}
\Rightarrow T_{n+1}\ne T_n$.

{\bf Proposition 5.} If $\la 2\ra$ is split, then for all $n$, $T_n'=T_n$.
In particular, $T'_n$ is the same for all ideal orbits
that do not belong to the sublattice $2^{n-1}\times 2^{n-1}$,
no matter what the ideal is.

{\bf Proof.\ \ }
Because $\la 2\ra$ is split, we have $\la 2\ra=P_1P_2$.
Let $T$ and $S$ be the smallest integers such that
$\lambda^T\equiv 1\mod{P_1^n}$ and
$\lambda^S\equiv 1\mod{P_2^n}$.
We prove that $T=S$.
First, we note that $P_1$ and $P_2$
are conjugate ideals, i.e., $P_1=P_2^*$.
We assume $T=S+R$ and $R\ge 0$.
Taking the conjugate of the congruence $\lambda^S\equiv 1\mod{P_2^n}$,
we obtain $\lambda^{*S}\equiv 1\mod{P_1^n}$, where $\lambda^*=\lambda^{-1}$.
Therefore,
$\lambda^S\lambda^{*S}\lambda^R\equiv \lambda^T\equiv 1\mod{P_1^n}$
$\Rightarrow \lambda^R\equiv 1\mod{P_1^n}$, i.e.,
there exists an integer $l\ge 0$ such that $R=lT$.
Because $T=S+lT$, we have $l=0 \Rightarrow T=S$.

Let $z$ belong to an ideal orbit of length $T'_n$
and $z\in P_2^k$, $z\not\in P_2^{k+1}$,
where $k\in\{1,2,\dots,n\}$.
Then $T'_n$ and $T_n$ are the smallest integers such that
$\lambda^{T'_n} \equiv 1\mod{P_1^nP_2^{n-k}}$
and $\lambda^{T_n} \equiv 1\mod{P_1^nP_2^n}$. Therefore,
$T'_n|T_n$.
On the other hand, $\lambda^{T'_n} \equiv 1\mod{P_1^n}
\Rightarrow \lambda^{T'_n} \equiv 1\mod{P_2^n}
\Rightarrow \lambda^{T'_n} \equiv 1\mod{P_1^nP_2^n}$,
i.e., $T_n|T'_n$. Therefore, $T_n'=T_n$.

{\bf Proposition 6.} If $\la 2\ra$ is ramified, then
for all $n$, either $T_n'=T_n$ or $T_n'=T_{n-1}$.

{\bf Proof.\ \ }
We have $\la 2\ra=P^2$.
We consider an orbit belonging to $P$. We now show that
the orbit period is either $T_n$ or $T_{n-1}$.
$$\left\{
\begin{array}{l}
\lambda^{T_{n-1}} \equiv 1\mod{\la 2^{n-1}\ra}, \\
\lambda^{T'_n} \equiv 1\mod{\la 2^{n-1}\ra P}, \\
\lambda^{T_n} \equiv 1\mod{\la 2^n\ra}. \\
\end{array}
\right.\Rightarrow T_{n-1}|T'_n|T_n.$$
Using Proposition 3, we complete the proof.

{\bf Proposition 7.} If $\la 2\ra$ is ramified, $n\ge3$ and
$T_n=2T_{n-1}$, then $T'_{n+1}= 2T'_n$.

{\bf Proof.\ \ }
Let $T=T_{n-1}$. Then we have
\begin{equation}
\left\{
\begin{array}{l}
\lambda^{T} \equiv 1\mod{A}, \\
\lambda^{T} \not\equiv 1\mod{AP}, \\
\end{array}
\right.
\label{lemma7}
\end{equation}
where $A=\la 2^{n-1}\ra$ for $T'_n=T_n$ and $A=\la 2^{n-1}\ra P$ for
$T'_n=T_{n-1}$.
In any case, $\la 2^{n-1}\ra |A$, $AP|\la 2^n\ra$.
It follows from~(\ref{lemma7}) that $\lambda^T=1+z$, where
$z\in A$, $z\not\in AP$. Hence, $\lambda^{2T}=1+2z+z^2$.
We note that $2z\in (\la 2\ra A)$, $2z\not\in (\la 2\ra AP)$,
and $z^2\in\la 2^{n+1}\ra \Rightarrow z^2\in (\la 2\ra AP)$
for $n\ge 3$.
Therefore,
\begin{equation}
\left\{
\begin{array}{l}
\lambda^{2T} \equiv 1\mod{\la 2\ra A}, \\
\lambda^{2T} \not\equiv 1\mod{\la 2\ra AP}. \\
\end{array}
\right.
\end{equation}

If $T'_n=T_n$, this means that
$T'_{n+1}\ne T_n \Rightarrow T'_{n+1}=T_{n+1}$.
In the case where $T'_n=T_{n-1}$, we have $T'_{n+1}=T_n$.
In any case, $T'_{n+1}=2T'_n$.

\section{Realizations and algorithms}
\label{RealizationSec}

\subsection{RNG realizations in C language and in inline assembler,
speed of realizations}

In this section, we present efficient algorithms for several
versions of the RNG introduced in Sec.~\ref{GeneratorSec}.
In particular, GS (cat map Generator, Simple version),
GR (cat map Generator, with Rotation),
GRI (cat map Generator, with Rotation, with Increased trace),
GM (cat map Generator, Modified version).
The parameters and characteristics for these generators
can be found in Table~\ref{RNGTable},
and the results of stringent statistical tests in Sec.~\ref{StatTestsSec}.
For comparison, both in Table~\ref{RNGTable} and
in Sec.~\ref{Batteries}, we also test the standard UNIX generators
\verb#rand()#, \verb#rand48()# and \verb#random()#
and the modern generators MT19937~\cite{MT}, MRG32k3a~\cite{CombinedLCG} and
LFSR113~\cite{LFSR113}
(see Sec.~\ref{Batteries} for details on them).

\begin{table}[hbt]
\caption{Characteristics and parameters for several versions of the
RNG based on the ensemble of cat maps (see Sec.~\ref{GeneratorSec})
and for other generators (last six entries).
Here ``CPU-time'' means the CPU time (in seconds) needed to generate
$10^8$ uniform random numbers on a 3.0 GHz Pentium 4 PC running Linux.
This parameter characterizes the speed of the generator.
Generators may be used only when the application needs not
more that $T$ random numbers, where $T$ is the RNG period.}
\begin{tabular}{|l|c|c|c|c|c|c|c|c|}
\hline
Generator & $g$ & $s$ & $k$ & $q$ & Rotation & SSE2 & Period & CPU-time\\
\hline
GS       & $2^{32}$   & $32$  & $3$  & $1$  & $-$ & $-$ & $3.2\cdot 10^9$      & $55.4$ \\
GS-SSE   & $2^{32}$   & $32$  & $3$  & $1$  & $-$ & $+$ & $3.2\cdot 10^9$      & $2.49$ \\
GR-SSE   & $2^{32}$   & $32$  & $3$  & $1$  & $+$ & $+$ & $3.2\cdot 10^9$      & $2.79$ \\
GSI-SSE  & $2^{32}$   & $32$  & $11$ & $1$  & $-$ & $+$ & $3.2\cdot 10^9$      & $3.66$ \\
GRI      & $2^{32}$   & $32$  & $11$ & $1$  & $+$ & $-$ & $3.2\cdot 10^9$      & $78.2$ \\
GRI-SSE  & $2^{32}$   & $32$  & $11$ & $1$  & $+$ & $+$ & $3.2\cdot 10^9$      & $4.03$ \\
GM19     & $2^{19}-1$ & $32$  & $6$  & $3$  & $+$ & $-$ & $2.7\cdot 10^{11}$   & $120.5$\\
GM19-SSE & $2^{19}-1$ & $32$  & $6$  & $3$  & $+$ & $+$ & $2.7\cdot 10^{11}$   & $6.11$ \\
GM31-SSE & $2^{31}-1$ & $32$  & $7$  & $11$ & $+$ & $+$ & $4.6\cdot 10^{18}$   & $8.86$ \\
\hline
RAND     & $-$        & $-$   & $-$  & $-$  & $-$ & $-$ & $2.1\cdot 10^9$      & $2.48$ \\
RAND48   & $-$        & $-$   & $-$  & $-$  & $-$ & $-$ & $2.8\cdot 10^{14}$   & $4.64$ \\
RANDOM   & $-$        & $-$   & $-$  & $-$  & $-$ & $-$ & $3.4\cdot 10^{10}$   & $1.88$ \\
MT19937  & $-$        & $-$   & $-$  & $-$  & $-$ & $-$ & $4.3\cdot 10^{6001}$ & $2.45$ \\
MRG32k3a & $-$        & $-$   & $-$  & $-$  & $-$ & $-$ & $3.1\cdot 10^{57}$   & $11.14$ \\
LFSR113  & $-$        & $-$   & $-$  & $-$  & $-$ & $-$ & $1.0\cdot 10^{34}$   & $2.98$ \\
\hline
\end{tabular}
\label{RNGTable}
\end{table}

Most of our generators are speeded up using Eq.~(\ref{Recurr})
instead of Eq.~(\ref{MatrixRecurr}).
Also, the Streaming SIMD Extensions 2 (SSE2) technology, introduced in
Intel Pentium 4 processors~\cite{Pentium4}, allows using 128-bit
XMM-registers to accelerate computations. A similar technique
was previously used for other generators~\cite{RS}. The SSE2 algorithms
for our generator are able to increase performance up to 23 times
as compared with usual algorithms (see Table~\ref{RNGTable}).

The algorithms for GS, GRI and GM19 are shown in Table~\ref{UsualAlgs}.
Table~\ref{EquivAlgs} illustrates the key ideas for speeding up cat-map
algorithms using SSE2. We use the GCC inline assembler syntax
for the SSE2 algorithms.
The action of the fast SSE2 algorithms shown in the left column are
equivalent to the action of the slow algorithms shown in the right column.

The complete realizations for all RNGs can be found in~\cite{AlgSite}.
GM31-SSE is the only algorithm here that exploits 64-bit
SSE-arithmetic for calculating Eq.~(\ref{Recurr}).
We must also note that the algorithms that exploit the SSE2 command set
work properly for Pentium processors starting from Pentium IV.
Therefore, some of our codes are not immediately portable.
Even the AMD's implementation of SSE2 is based on a slightly
different command set.

\begin{table}[hbt]
\caption{Codes in ANSI C language for the generators GS, GRI and GM19.}
\newsavebox{\FirstAlgBox}
\newsavebox{\SecondAlgBox}
\newsavebox{\ThirdAlgBox}
\newsavebox{\FourthAlgBox}

\begin{lrbox}{\FirstAlgBox}
\begin{minipage}{7.3cm}
\begin{verbatim}

const unsigned long halfg=2147483648;
unsigned long x[32],y[32]; char rotate;

//----------- Generator GS -----------------

unsigned long GS(){
  unsigned long i,output=0,bit=1;
  for(i=0;i<32;i++){
    x[i]=x[i]+y[i];
    y[i]=x[i]+y[i];
  }
  for(i=0;i<32;i++){
    output+=((x[i]<halfg)?0:bit); bit*=2;}
  return output;
}

//----------- Generator GRI ----------------

unsigned long GRI(){
  unsigned long i,oldx,oldy,output=0,bit=1;
  oldx=x[31]; oldy=y[31];
  for(i=31;i>0;i--){
    x[i]=4*x[i-1]+9*y[i-1];
    y[i]=3*x[i-1]+7*y[i-1];
  };
  x[0]=4*oldx+9*oldy; y[0]=3*oldx+7*oldy;
  for(i=0;i<32;i++){
    output+=((x[i]<halfg)?0:bit); bit*=2;}
  rotate++; return output;
}

\end{verbatim}
\end{minipage}
\end{lrbox}

\begin{lrbox}{\SecondAlgBox}
\begin{minipage}{10.2cm}
\begin{verbatim}

//-------------------- Generator GM19 -----------------------

const unsigned long k=14;
const unsigned long q=15;
const unsigned long g=524287;
const unsigned long qg=7864305;
const unsigned long halfg=262143;
unsigned long x[2][32]; char new,rotate;

unsigned long GM19(){
  unsigned long i,output=0,bit=1;
  char old=1-new;
  for(i=0;i<32;i++)
    x[old][i]=(qg+k*x[new][i]-q*x[old][i])%g;
  for(i=0;i<32;i++){
    output+=((state->x[old][(256+i-rotate)%32]<halfg)?0:bit);
    bit*=2;
  }
  new=old; rotate++; return output;
}

\end{verbatim}
\end{minipage}
\end{lrbox}

\begin{tabular}{|c|c|}
\hline
\usebox{\FirstAlgBox} & \usebox{\SecondAlgBox}\\
\hline
\end{tabular}
\label{UsualAlgs}
\end{table}

\begin{table}
\caption{Equivalent realizations for several algorithms with
inline assembler code for Pentium 4 processor (left column) and
ANSI C language (right column).
First row presents the main part of the GRI algorithm.
Second row presents the packing $16$ high bits of $16$ integers
into one integer.
These or similar equivalences are used in constructing the SSE2 algorithms
for any of the discussed RNGs~\cite{AlgSite}.}
\begin{lrbox}{\FirstAlgBox}
\begin{minipage}{6cm}
\begin{verbatim}

  unsigned long x[4],y[4];

  [.......]

  asm("movaps (%0),%%xmm0\n" \
      "movaps (%1),%%xmm1\n" \
      "paddd  %%xmm1,%%xmm0\n" \
      "paddd  %%xmm1,%%xmm0\n" \
      "movaps %%xmm0,%%xmm2\n" \
      "pslld  $2,%%xmm0\n" \
      "paddd  %%xmm1,%%xmm0\n" \
      "movaps %%xmm0,(%0)\n" \
      "psubd  %%xmm2,%%xmm0\n" \
      "movaps %%xmm0,(%1)\n" \
      ""::"r"(x),"r"(y));

\end{verbatim}
\end{minipage}
\end{lrbox}

\begin{lrbox}{\SecondAlgBox}
\begin{minipage}{6cm}
\begin{verbatim}

  unsigned long i,newx[4],x[4],y[4];

  [.......]

  for(i=0;i<4;i++){
    newx[i]=4*x[i]+9*y[i];
    y[i]=3*x[i]+7*y[i];
    x[i]=newx[i];
  }

\end{verbatim}
\end{minipage}
\end{lrbox}

\begin{lrbox}{\ThirdAlgBox}
\begin{minipage}{6cm}
\begin{verbatim}

unsigned long x[16],output;

  [.......]

  asm("movaps (%1),%%xmm0\n" \
      "movaps 16(%1),%%xmm1\n" \
      "movaps 32(%1),%%xmm2\n" \
      "movaps 48(%1),%%xmm3\n" \
      "psrld  $31,%%xmm0\n" \
      "psrld  $31,%%xmm1\n" \
      "psrld  $31,%%xmm2\n" \
      "psrld  $31,%%xmm3\n" \
      "packssdw %%xmm1,%%xmm0\n" \
      "packssdw %%xmm3,%%xmm2\n" \
      "packsswb %%xmm2,%%xmm0\n" \
      "psllw  $7,%%xmm0\n" \
      "pmovmskb %%xmm0,%0\n" \
      "":"=r"(output):"r"(x));

\end{verbatim}
\end{minipage}
\end{lrbox}

\begin{lrbox}{\FourthAlgBox}
\begin{minipage}{6.5cm}
\begin{verbatim}

const unsigned long halfg=2147483648;
unsigned long x[16],i,output=0,bit=1;

  [.......]

for(i=0;i<16;i++){
  output+=((x[i]<halfg)?0:bit;
  bit*=2;
}

\end{verbatim}
\end{minipage}
\end{lrbox}

\begin{tabular}{|c|c|}
\hline
\usebox{\FirstAlgBox} & \usebox{\SecondAlgBox}\\
\hline
\hline
\usebox{\ThirdAlgBox} & \usebox{\FourthAlgBox}\\
\hline
\end{tabular}
\label{EquivAlgs}
\end{table}

\subsection{Initialization of generators}

The proper initialization is very important for a good generator.

For the generators GS, GS-SSE, GR-SSE, GRI, GSI-SSE and GRI-SSE we use
the following initialization method:
\begin{itemize}
\item Norms of all points should be different modulo $256$. In particular,
this guarantees that the initial points ${x_i^{(0)}\choose y_i^{(0)}}$,
$i=0,1,\dots,(s-1)$ belong to different orbits of the cat map,
and that none of the symmetries may convert one orbit to another
(see Appendix~\ref{NormDiscussion}).
\item At least one point should belong to a free orbit, i.e., at least
one of the coordinates $x$ or $y$ should be an odd number.
This guarantees that the period length is not smaller than $T_m$
(see Appendix~\ref{RNGPeriod}).
\end{itemize}

We choose the parameters $k$ and $q$ for the generators GM19 and GM31
such that the polynomial $f(x)=x^2{-}kx{+}q$ is primitive modulo
$p$, where $p=2^{19}{-}1$ for GM19 and $p=2^{31}{-}1$ for GM31.
Therefore, the actual period of the generator is $p^2{-}1$.

To construct the initialization method for GM19 and GM31,
we use the ``jumping ahead'' property, the possibility to
skip over terms of the generator.
In other words, we utilize an easy algorithm to calculate $x_n$
quickly from $x_0$ and $x_1$, for any large $n$.
We choose the following initial conditions: $x_i^{(0)}=x_{i*A}$,
$x_i^{(1)}=x_{i*A+1}, i=0,1,\dots,31$. Here we follow the
notation in Sec.~\ref{GeneratorSec} and $A$ is a
value of the order of $(p^2{-}1)/32$. We recommend to choose $A$
randomly; at least $A$ should not be chosen very close to the
divisor of $p^2-1$ or to a large power of $2$.
We recommend using less than $A$ random numbers in applications
that use GM19 and GM31. The values of $A$ are approximately $32$
times smaller than the periods in Table~\ref{RNGTable}.

The initialization routines for all generators can also be found
in~\cite{AlgSite}.

\end{document}